\documentclass[aps, pre, twocolumn, nofootinbib]{revtex4-1}
\usepackage{graphicx}
\begin{document}
%%%%%%%%%%%%%%%%%%%%%%
\title{A Model of Electrowetting, Reversed Electrowetting and Contact Angle Saturation}
\author{Dan Klarman, David Andelman}
\email{andelman@post.tau.ac.il}
\affiliation{Raymond \& Beverly Sackler School of  Physics and Astronomy,
Tel-Aviv University, Ramat Aviv 69978, Tel Aviv, Israel}
\author{Michael Urbakh}
\affiliation{Raymond \& Beverly Sackler School of Chemistry,
Tel-Aviv University, Ramat Aviv 69978, Tel Aviv, Israel}
\date{April 18, 2011}

%%%%%%%%%%%%%%%%%%%%%%%%%
\begin{abstract}
%%%%%%%%%%%%%%%%%%%%%%%%&
While electrowetting has many applications, it is limited at large  voltages by \emph{contact angle saturation}  --- a phenomenon that is still  not well understood. We propose a generalized approach for electrowetting that, among other results, can shed new light on contact angle saturation. The model assumes the existence of a minimum (with respect to the contact angle) in the electric energy and accounts for a quadratic voltage dependence $\sim U^2$  in the low-voltage limit,  compatible with the Young-Lippmann formula, and a $\sim U^{-2}$ saturation at the high-voltage limit. Another prediction is the surprising possibility of a reversed electrowetting regime, in which the contact angle increases with applied voltage. By explicitly taking into account the effect of the counter-electrode, our model is shown to be applicable to several AC and DC experimental electrowetting-on-dielectric (EWOD) setups. Several features seen in experiments compare favorably with our results. Furthermore, the AC frequency dependence of EWOD agrees quantitatively with our predictions. Our numerical results are complemented with  simple analytical expressions for the saturation angle in two practical limits.
\end{abstract}
%%%%%%%%%%%%%%%
\maketitle
%\newpage

%%%%%%%%%%%%%%%%%%%%%%%%%%%%
\section{Introduction}
\begin{figure}[h]
\begin{center}
%\vspace{0.7cm} \centerline{\resizebox{0.6\textwidth}{!}
\includegraphics[width=8cm]{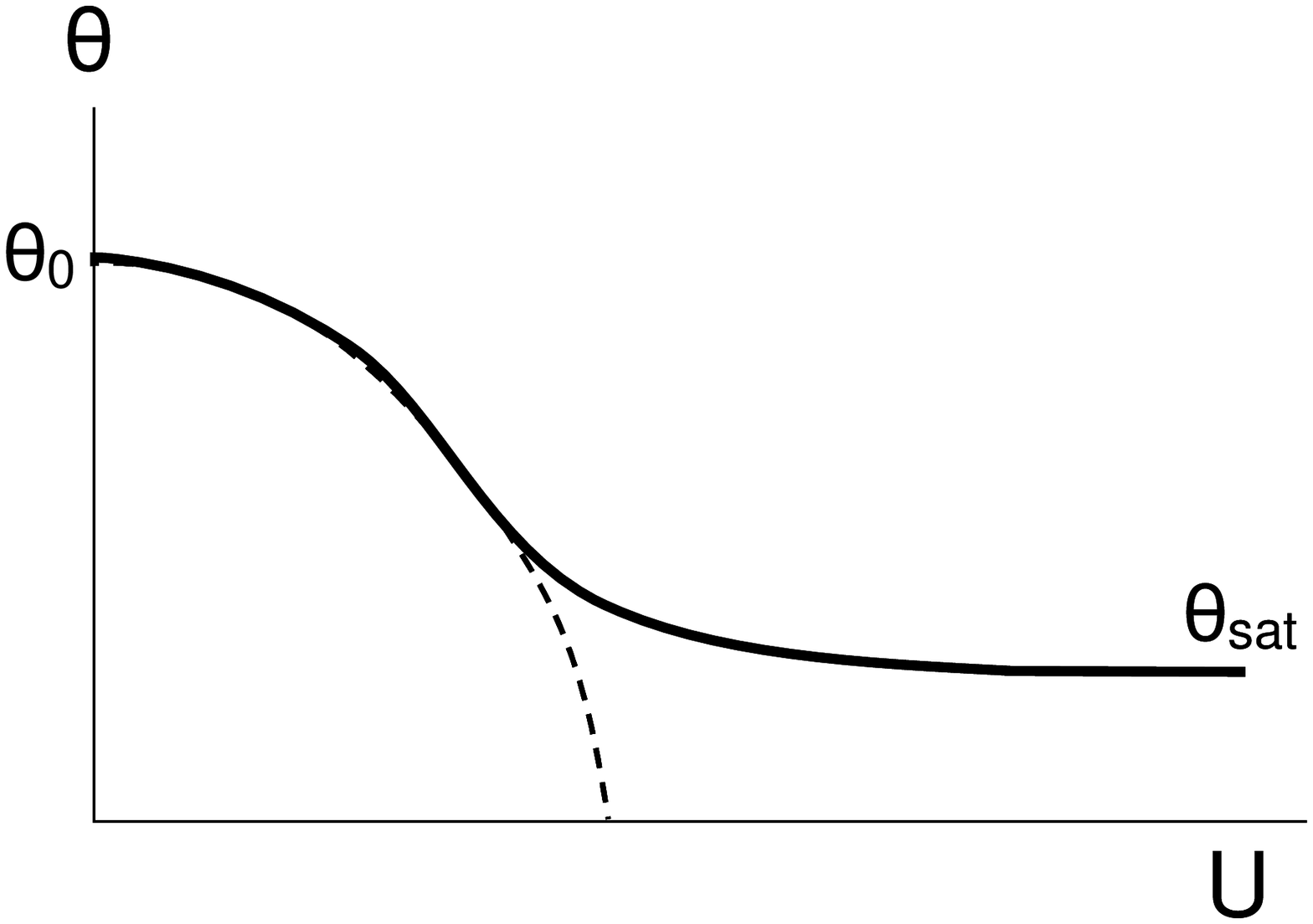}
\caption{\label{fig:exp_sketch}\textsf{A schematic plot of electrowetting contact angle, $\theta(U)$.
For zero voltage, the contact angle is the same as the Young angle, $\theta_{0}$.
At low applied voltages $U$, the contact angle follows the
Young-Lippmann formula, eq~\ref{eq:Y-L_formula}, (dashed line) $\cos\theta(U)-\cos\theta_0\simeq \cos\theta_{\rm YL}(U)-\cos\theta_0\sim U^2$,
but for higher voltages the contact angle gradually
deviates from the Young-Lippmann behavior and saturates
towards some finite value, $\theta_{\rm sat}$. }}
\end{center}
\end{figure}

%%%%%%%%%%%%%%%%%%%%%%%%%%%%
The term \emph{Electrowetting}, in its broadest sense, refers to techniques
by which one can control the apparent wettability (characterized by the contact angle) of liquids, by applying an external electric
potential~\cite{Mimmena_1980,Beni_1980,Gorman_1995,Sondag-Huethorst_1994,Vallet_Berge_Vovelle_1996,Welters_Fokkink_1998,Ionization_1999,FreqEW_2007,ITIES1_2007,ITIES2_2006,Jones_etal1_freq_2003}.
While it has numerous applications~\cite{B2app_2005,Manipulation_1999,Romi_2008,Electrospray_Slata_2005,Microfluidics_Pollack_2000,Microfluidics_Troian_2005,EWdisplays_Hayes_2006,Berge_2000,InkJetEW_2006},
electrowetting is known to be limited by the so-called \emph{contact angle saturation} (CAS)~\cite{Vallet_Berge_Vovelle_1996,Ionization_1999,ElectrostaticLimitations_Adamiak_2006,finiteR_2003,Irreversibility_2009,ZeroCapillary_2005,Berge_2001,ChargeTrapping_1999,lowV_2006} as depicted in figure~\ref{fig:exp_sketch}.
As the term indicates, an electric potential can incur a change in the contact angle, but only to a certain extent. Further voltage increase has  no additional effect on the contact angle. This behavior is not accounted for by the
standard model of electrowetting, and its
origin still remains a point of controversy~\cite{Vallet_Berge_Vovelle_1996,Welters_Fokkink_1998,Ionization_1999,B2app_2005,ElectrostaticLimitations_Adamiak_2006,finiteR_2003,Irreversibility_2009,ZeroCapillary_2005,Berge_2001,ChargeTrapping_1999,lowV_2006,Polarity_2006,Jones_etal3_freq_2005,Jones_etal2_freq_2004,papathanasiou_2007,Fontelos_2009,papathanasiou_2005}.

%%%%%%%%%%%%%%%%%%%%%%%%%%%%%%%%%%%%%%%%%%%%%%%%%%%%%%%%%%%%%%%%%%%%%%%%%%%%%%
%fig1
%%%%%%%%%%%%%%%%%%%%%%%%%%%%%%%%%%%%%%%%%%%%%%%%%%%%%%%%%%%%%%%%%%%%%%%%%%%%%%
%%%%%%%%%%%%%%%%%%%%%%%%%%%%%%%%%%%%%%%%%%%%%%%%%%%%%%%%%%%%%%%%%%%%%%%%%%%%%%

When a small drop of liquid is placed on top of a solid surface it assumes the shape of a spherical cap~\cite{Young_1805}. The contact angle $\theta_{0}$
between the drop and the surface, given by the Young formula~\cite{DeGennesBook,DeGennesReview,Young_1805}
\begin{equation}
\cos\theta_{0}=\frac{\gamma_{\rm sa}-\gamma_{\rm sl}}{\gamma_{\rm la}}\label{eq:Young}
\end{equation}
depends on the three interfacial tensions:   solid/air  $\gamma_{\rm sa}$,  solid/liquid  $\gamma_{\rm sl}$ and
liquid/air $\gamma_{\rm la}$, where the air phase can be replaced by another immiscible fluid~\cite{Ionization_1999,Microfluidics_Troian_2005,Berge_2000}.

\begin{figure}[h]
\begin{center}
\includegraphics[width=8cm]{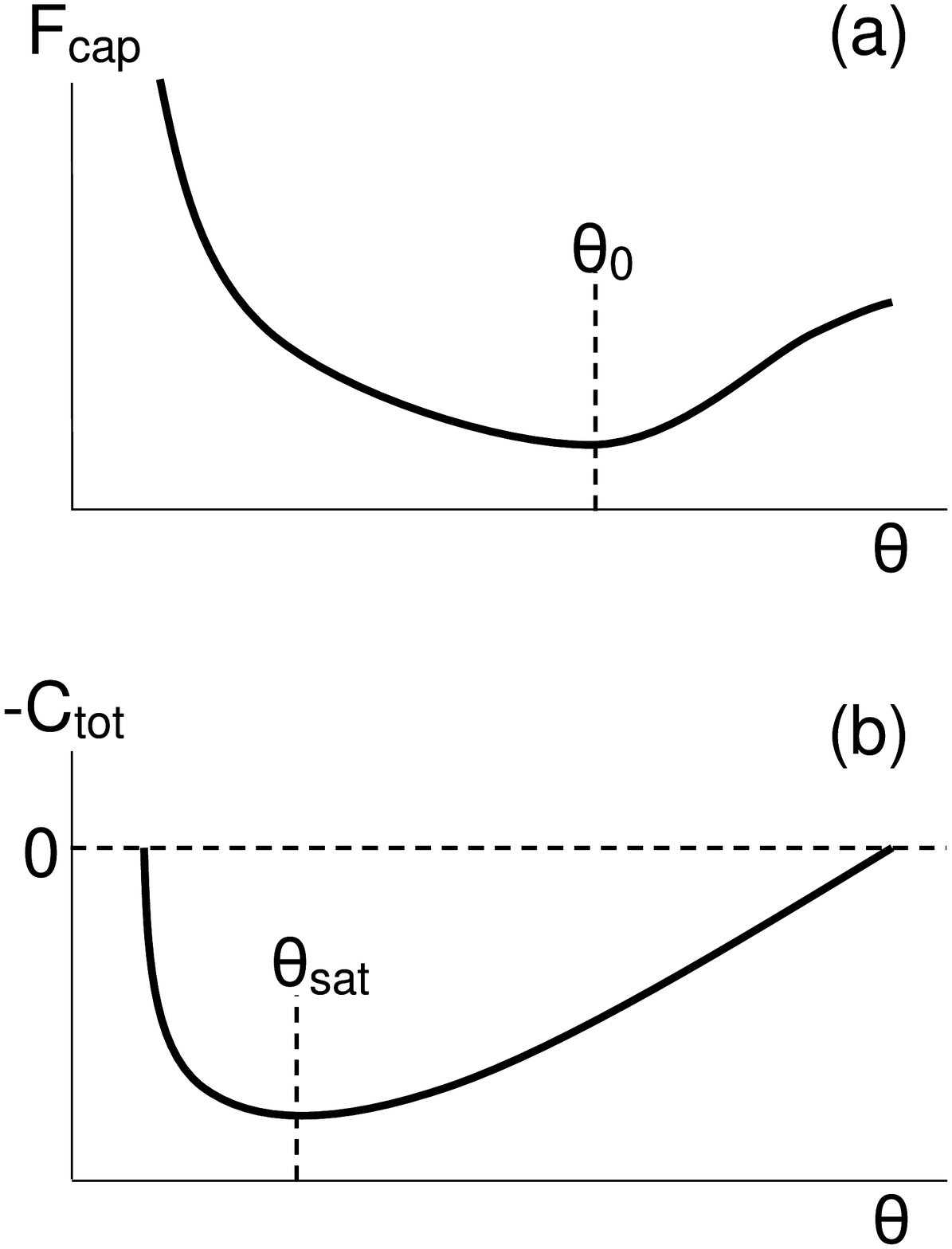}
\caption{\label{fig:min_energy_sketch}\textsf{Schematic plots of (a) the capillary free energy $F_{\rm cap}$
for a spherical drop of fixed volume and (b) a hypothetical (negative) total capacitance, $-C_{\rm tot}$, as function
of contact angle $\theta$  in electrowetting setup.  The minimum
of the capillary term $F_{\rm cap}$ occurs at the Young angle $\theta_{0}$, while
the electric term $F_{\rm el}=-\frac{1}{2}C_{\rm tot}U^{2}$
(or equivalently of $-C_{\rm tot}$) is assumed to have a minimum at a finite saturation angle, $\theta_{\rm sat}$.}}
\end{center}
\end{figure}

The Young formula, eq~\ref{eq:Young}, can  be obtained by minimizing
the capillary free energy with a fixed volume constraint~\cite{DeGennesReview}
\begin{equation}
F_{\rm cap}(\theta)=A_{\rm sa}\gamma_{\rm sa}+
A_{\rm sl}\gamma_{\rm sl}+A_{\rm la}\gamma_{\rm la}-V\Delta P\label{eq:YoungModel}
\end{equation}
where $A_{ij}$ are the interface areas between the $i$ and $j$ phases, $i,j=$~a (air), l (liquid), and s (solid);
the drop volume is $V$ and  the pressure difference across the liquid/air interface is $\Delta P$.
For partial wetting, $\gamma_{\rm sa}<\gamma_{\rm sl}+\gamma_{\rm la}$,
 the capillary free energy $F_{\rm cap}$ has a minimum at the Young angle,
$\theta_{0}$ [figure~\ref{fig:min_energy_sketch}(a)].

%%%%%%%%%%%%%%%%%%%%%%%%%%%%%%%%%%%%%%%%%%%%%%%%%%%%%%%%%%%%%%%%%%%%%%%%%%%%%%
%fig2
%%%%%%%%%%%%%%%%%%%%%%%%%%%%%%%%%%%%%%%%%%%%%%%%%%%%%%%%%%%%%%%%%%%%%%%%%%%%%%
%\newpage{}
%%%%%%%%%%%%%%%%%%%%%%%%%%%%%%%%%%%%%%%%%%%%%%%%%%%%%%%%%%%%%%%%%%%%%%%%%%%%%%

\begin{figure}[h]
\begin{center}
\includegraphics[width=5cm]{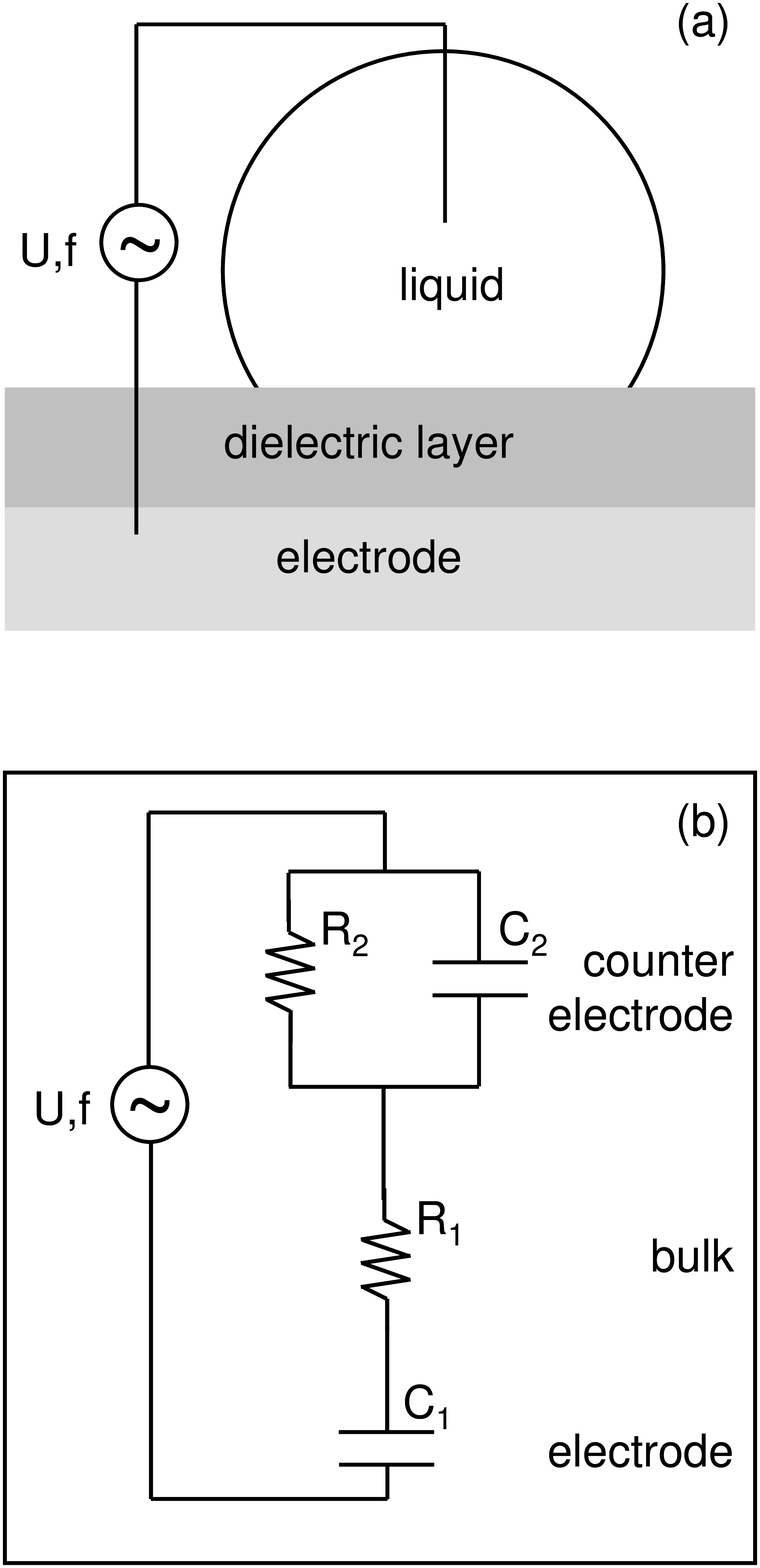}
\caption{\label{fig:AC_equiv_circ}\textsf{(a) A sketch of an EWOD setup with AC voltage $U$ of frequency $f$, and (b) its
equivalent AC circuit. The two parallel-plate capacitors
of capacitances $C_{1}$ and $C_{2}$ and  areas $A_1$ and $A_2$ represent
the substrate/liquid and the counter-electrode/liquid interfaces, respectively. The bulk liquid drop is
represented by a resistor $R_{1}$ through which the two capacitors
are charged and discharged. The charging time through the resistor is equal to the
build-up time $\tau_{b}$ of the double-layers at the two interfaces. The discharge resistor $R_{2}$
represents the Faradic charge transfer
processes that relax the electric double-layer near the counter-electrode
with relaxation time $\tau_{r}$. Such mechanism is
prevented at the substrate electrode because of its dielectric coating.
}}
\end{center}
\end{figure}

The contact angle $\theta$ can be  varied from its initial value $\theta_{0}$
by applying an external voltage of several volts to several hundreds of volts
across the liquid drop.
A commonly used electrowetting setup
developed by Rinkel et al~\cite{Mimmena_1980} and
later perfected by Vallet {\rm et al}~\cite{Vallet_Berge_Vovelle_1996} is called
electrowetting-on-dielectric (EWOD).
The apparatus, roughly sketched in figure~\ref{fig:AC_equiv_circ}(a), includes a flat electrode as a
substrate, which is  coated with a thin dielectric layer
(tens of nanometers to several micrometers thick), whose purpose
is to prevent Faradic  charge exchange ({\it i.e.},
electrochemical reactions) at the electrode. It is common that this dielectric layer is then topped
with an even thinner hydrophobic ({\it i.e.}, Teflon) layer in order to control its surface tension.
A drop of  ionic solution is placed atop
the coated electrode  and a thin counter-electrode (usually a bare platinum fiber) is inserted
into the drop from above. The drop is surrounded by air or by another immiscible  dielectric liquid.
Applying a voltage across the drop can cause a  large change, of several tens of degrees, in
the contact angle.

%%%%%%%%%%%%%%%%%%%%%%%%%%%%%%%%%%%%%%%%%%%%%%%%%%%%%%%%%%%%%%%%%%%%%%%%%%%%%%
%fig3
%%%%%%%%%%%%%%%%%%%%%%%%%%%%%%%%%%%%%%%%%%%%%%%%%%%%%%%%%%%%%%%%%%%%%%%%%%%%%%
%%%%%%%%%%%%%%%%%%%%%%%%%%%%%%%%%%%%%%%%%%%%%%%%%%%%%%%%%%%%%%%%%%%%%%%%%%%%%%

As reviewed in Ref~\citenum{B2app_2005}, a simple relation between the contact angle and the applied
voltage can be derived. When an external voltage $U$ is  applied an electric double-layer is formed at the
liquid/substrate interface. The total free energy $F_{\rm tot}$
has two contributions: a capillary term $F_{\rm cap}$, defined~\cite{ft3} in
eq~\ref{eq:YoungModel}, and an electric term, $F_{\rm el}$ that depends on $\theta$, $U$ and other system parameters.
\begin{equation}
F_{\rm tot}(\theta,U)=F_{\rm cap}(\theta)+F_{\rm el}(\theta,U)
\label{eq:Ftot}
\end{equation}
Within the standard model of electrowetting (under external voltage control) the electric term is evaluated as
\begin{equation}
F_{\rm el}= -\frac{1}{2}C_{\rm ld}U^{2}
\label{eq:Ftot1}
\end{equation}
where $C_{\rm ld}$ is the capacitance
of the liquid/substrate interface, which is modeled as a parallel-plate capacitor:
\begin{equation}
C_{\rm ld}\simeq \varepsilon_{0}A_{\rm ld}\left(\frac{d}{\varepsilon_d}+\frac{\lambda_D}{\varepsilon_l}\right)^{-1}
\simeq \frac{\varepsilon_{0}\varepsilon_{d}}{d}A_{\rm ld}
\label{eq:Cld}
\end{equation}
where $A_{\rm ld}$ is the substrate area that is covered by the
liquid drop, $d$ is the width of the dielectric layer and $\varepsilon_d$ and $\varepsilon_l$ are the dielectric
constants of the dielectric coating and liquid drop, respectively. The Debye screening length, $\lambda_D$, is the width of the electric double-layer.  In most cases, $d/\varepsilon_d \gg \lambda_D/\varepsilon_l$, and the second approximation in eq~\ref{eq:Cld} can be justified.
The parallel-plate capacitor
model has been shown~\cite{ITIES2_2006} to be valid as long as fringe fields are
negligible ($d\ll A_{\rm ld}^{1/2}$) and the drop
of volume $V$ is not too small for double-layers to be created ($\lambda_D\ll V^{1/3}$).
An implicit assumption made in  eq~\ref{eq:Ftot1} is that the counter-electrode
does not contribute to the total capacitance.

In an electrowetting setup, the contact angle $\theta(U)$ depends on the
external voltage $U$. By substituting  eq~\ref{eq:Cld} into eq~\ref{eq:Ftot1}
and minimizing $F_{\rm tot}$ of
eq~\ref{eq:Ftot} with a fixed volume constraint,
the Young-Lippmann  formula~\cite{B2app_2005} for the contact angle $\theta_{\rm YL}(U)$ is obtained
\begin{equation}
\cos\theta_{\rm YL}(U)=\cos\theta_{0}+
g^{-1}U^{2}
\label{eq:Y-L_formula}
\end{equation}
where $g^{-1}\equiv \varepsilon_{0}\varepsilon_{d}/(2\gamma_{\rm la}{d})$.
It is a common practice to extend this
DC voltage model to AC setups by using the rms voltage  in eq~\ref{eq:Y-L_formula}, $U^2\to U^2_{\rm rms}$. Note that
similar results can be obtained
using force balance at the three-phase contact line~\cite{Kang2002}.

Experiments~\cite{ElectrostaticLimitations_Adamiak_2006,Jones_etal3_freq_2005,Jones_etal2_freq_2004,B2app_2005,papathanasiou_2007,ZeroCapillary_2005,Irreversibility_2009,finiteR_2003,Vallet_Berge_Vovelle_1996,Ionization_1999,Welters_Fokkink_1998}
have shown that the $\sim U^2$ behavior predicted by the Young-Lippmann formula is indeed found for a range of low applied
voltages, but the pre-factor of the $U^2$ term, eq~\ref{eq:Y-L_formula},
does \emph{not} usually match the experimental data. For larger values of $U$,  a deviation from the $U^2$ behavior is observed
and a saturation in the contact angle, $\theta(U)\to\theta_{\rm sat}$, is reached gradually,
as is schematically sketched in figure~\ref{fig:exp_sketch}.
In addition, it is convenient to define a characteristic value of the cross-over voltage $U^*$ by requiring
that $\theta_{\rm YL}(U^*)=\theta_{\rm sat}$ in eq~\ref{eq:Y-L_formula}:
\begin{equation}
\label{eq:U_cross-over}
\left(U^*\right)^2= g(\cos\theta_{\rm sat}-\cos\theta_{0})
\end{equation}

Over the last decades, several models have been presented in an attempt to explain CAS~\cite{B2app_2005}. Most of these
models~\cite{Ionization_1999,Fontelos_2009,papathanasiou_2005,papathanasiou_2007,finiteR_2003,ChargeTrapping_1999}
are based on specific leakage mechanisms. Others, as in Ref~\citenum{ZeroCapillary_2005},
proposed heuristic arguments in order to predict CAS in electrowetting
systems without relying on a specific mechanism.

Considering that the origin of CAS is not well understood from general principles,
the objective of the present work is to offer a different approach to CAS, and to electrowetting in
general. In the following section we consider the general circumstances in which CAS can occur
intrinsically (without leakage). We present a low-voltage limit compatible with the Young-Lippmann
quadratic voltage dependence and a high-voltage limit in which  CAS is obtained.
Furthermore, we identify a  possibility for a novel electrowetting regime
we call {\it reversed electrowetting}. In section III we present an application of this approach
to EWOD experimental setups using a  geometry-dependent model, and
use AC circuit analysis to calculate the free energy.
Section IV is dedicated to showing several numerical and analytical results and their compatibility
with experiments.
We conclude in section V with a summary along with some further
discussion and an outlook on future research.

%%%%%%%%%%%%%%%%%%%%%%%%%%%%%%%%
\section{A Generalized Model of Electrowetting}
%%%%%%%%%%%%%%%%%%%%%%%%%%%%%%%%

%%%%%%%%%%%%%%%%%%%%%%%%%%%%%%
\subsection{Generalized free energy and contact angle saturation}
%%%%%%%%%%%%%%%%%%%%%%%%%%%%%%%
Our starting point is eq~\ref{eq:Ftot} above. Assuming that all the electric energy is
 stored via charge separation, it can be written in terms of the total capacitance,
$F_{\rm el}=-\frac{1}{2}C_{\rm tot}U^2$.
The total {\it electrocapillary} free energy $F_{\rm tot}$ is now written as%~\cite{ft2}:
\begin{equation}
F_{\rm tot}(\theta,U)=F_{\rm cap}(\theta)-
\frac{1}{2}C_{\rm tot}(\theta)U^{2}\label{eq:generalized_freeE}
\end{equation}
where all capillary contributions are included in $F_{\rm cap}$.
As $F_{\rm cap}$ is independent of
$U$, the relative magnitude of the two terms in eq~\ref{eq:generalized_freeE}
is controlled by the negative $U^2$ dependence of the $F_{\rm el}$ term.

Physical insight can be gained from eq~\ref{eq:generalized_freeE} by making different assumptions
regarding the behavior of $C_{\rm tot}(\theta)$.
Particularly, interesting results follow from the assumption that $C_{\rm tot}(\theta)$
has a maximum at some finite angle (see figure~\ref{fig:min_energy_sketch}(b)), as will be shown below to be the case
for EWOD setups. Our model therefore differs from previous models that took $C_{\rm tot}$
to be equal to $C_{\rm ld}$, (eq~\ref{eq:Ftot1}), which yields a monotonically decreasing $F_{\rm el}$.
For reasons to be immediately apparent, we denote
the angle where $C_{\rm tot}(\theta)$ has a maximum (or, equivalently, $F_{\rm el}$ has a minimum) as $\theta_{\rm sat}$.
This angle, in general, is different from the Young angle, $\theta_{0}$, which minimizes
the capillary term, $F_{\rm cap}$.

\begin{figure}[h]
\begin{center}
\includegraphics[width=6cm]{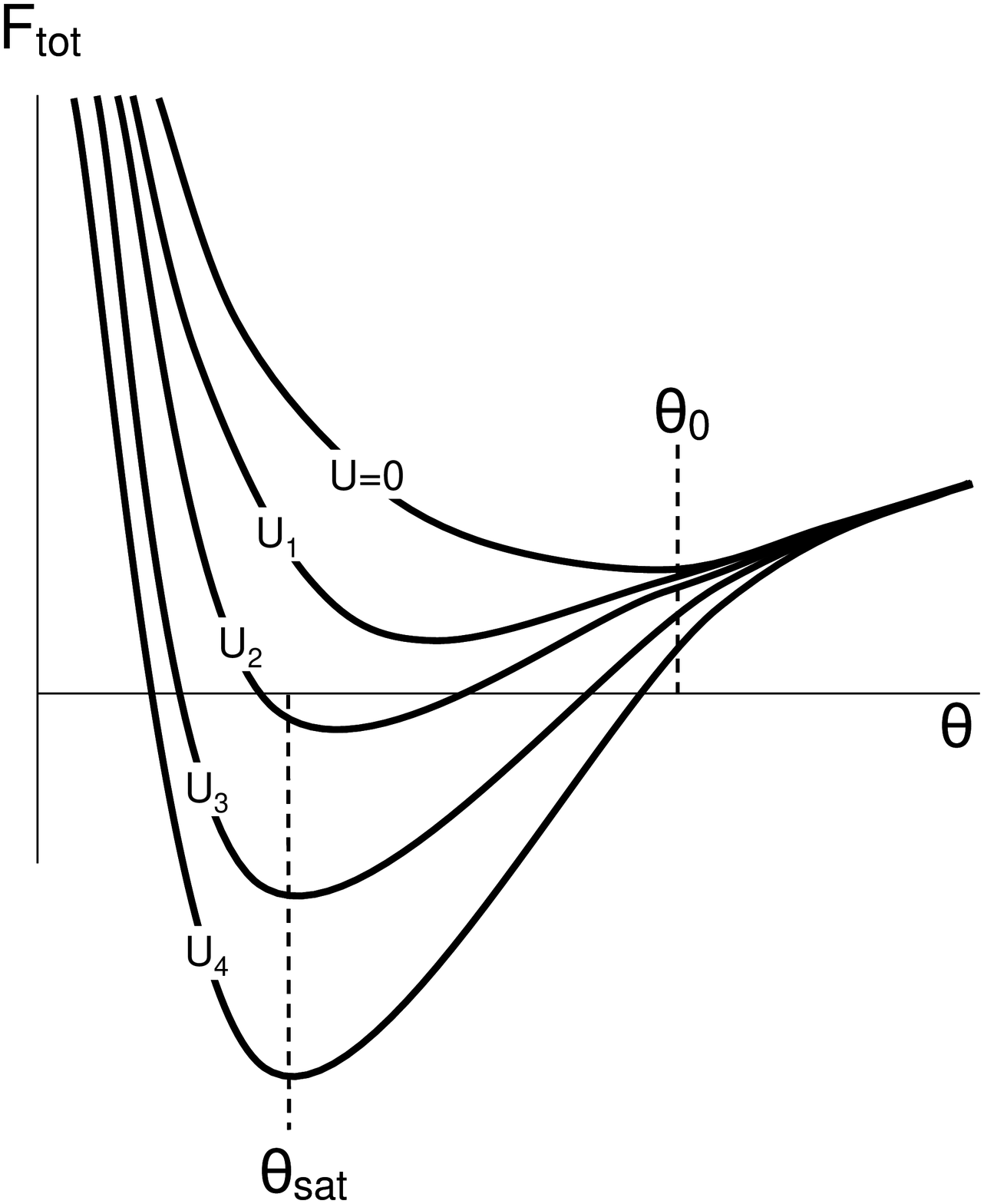}
\caption{\label{fig:saturation_sketch}\textsf{A schematic plot of the total free energy
$F_{\rm tot}(\theta,U)=
F_{\rm cap}(\theta)+F_{\rm el}(\theta,U)$, for a sequence of five applied voltages:
$U_4> U_3>U_2>U_1>U=0$.
Since the two energy terms are concave in the accessible range of $\theta$,
their sum is also concave and has
a minimum. Increasing $U$ from zero to a large $U_{4}$
causes a gradual shift of the minimum from $\theta_{0}$
towards $\theta_{\rm sat}$. Note that it cannot cross
beyond $\theta_{\rm sat}$ regardless of how high the voltage $U$
is  because $\theta_{\rm sat}$
is the minimum of the dominating $F_{\rm el}$ term. }}
\end{center}
\end{figure}

We now show how the existence of a global electric free-energy minimum at a finite contact angle yields CAS. With no applied voltage ($U=0$), $F_{\rm tot}=F_{\rm cap}$ and the system adheres to the Young
angle, $\theta=\theta_{0}$, (figure~\ref{fig:min_energy_sketch}(a)). Similarly, when the applied voltage
$U$ is  very large,
the free energy is  dominated by the electric term, $|F_{\rm el}|\propto U^2\gg F_{\rm cap}$,
and the system tends towards $\theta_{\rm sat}$, which minimizes $F_{\rm el}$ (or, equivalently,
maximizes $C_{\rm tot}$, figure~\ref{fig:min_energy_sketch}(b)). Now, if the two contributions
are concave for the accessible range of $\theta$,
then the minimum of $F_{\rm tot}$ shifts smoothly from $\theta_0$ towards $\theta_{\rm sat}$ as $U$
is increased from zero to
an arbitrary large value,
as is schematically illustrated in figure~\ref{fig:saturation_sketch}.
This description is consistent  with CAS and implies that the
saturation angle found in experiments
can be identified with our definition of $\theta_{\rm sat}$.

Below we analyze more quantitatively the consequences of such a global electric minimum at
the low- and high-voltage limits. In the former, we show a $\sim U^2$ variation of the
contact angle with a pre-factor that can match the Young-Lippmann formula or be different
from it. In the latter, an asymptotic $\sim U^{-2}$ approach to $\theta_{\rm sat}$ is found.

%%%%%%%%%%%%%%%%%%%%%%%%%%%%%%%%%%%%%%%%%%%%%%%%%%%%%%%%%%%%%%%%%%%%%%%%%%%%%%
%fig4
%%%%%%%%%%%%%%%%%%%%%%%%%%%%%%%%%%%%%%%%%%%%%%%%%%%%%%%%%%%%%%%%%%%%%%%%%%%%%%
%%%%%%%%%%%%%%%%%%%%%%%%%%%%%%%%%%%%%%%%%%%%%%%%%%%%%%%%%%%%%%%%%%%%%%%%%%%%%%

%%%%%%%%%%%%%%%%%%%%%%%%%%%%%%
\subsection{The Low Voltage Limit}
%%%%%%%%%%%%%%%%%%%%%%%%%%%%%%
For $U\rightarrow0$, the minimum of $F_{\rm tot}$ occurs
close to $\theta_0$.
Expanding this minimum condition
$F'_{\rm tot}(\theta)=0$ to first order in $\delta \theta=\theta(U)-\theta_0$,
while recalling that $F'_{\rm cap}(\theta_0)=0$, we obtain
\begin{equation}
 F''_{\rm cap}\left(\theta_0\right)\delta\theta-\frac{1}{2}U^{2}\left[C'_{\rm tot}\left(\theta_0\right)+
 C''_{\rm tot}\left(\theta_0\right)\delta\theta\right]  \simeq  0
\end{equation}
yielding,
\begin{equation}
\delta\theta\simeq \frac{1}{2}\frac{C'_{\rm tot}\left(\theta_0\right)}{F''_{\rm cap}
\left(\theta_0\right)-\frac{1}{2}U^{2}C''_{\rm tot}\left(\theta_0\right)}U^{2}
\end{equation}
and to leading order in $U^2$ one has
\begin{equation}
\label{eq:lowU}
\theta(U)\simeq \theta_0+\frac{1}{2}\frac{C'_{\rm tot}\left(\theta_0\right)}
{F''_{\rm cap}\left(\theta_0\right)}U^{2}
\end{equation}
and equivalently
\begin{equation}
\cos\theta(U)\simeq \cos\theta_0-\frac{1}{2}\frac{C'_{\rm tot}\left(\theta_0\right)\sin\theta_0}
{F''_{\rm cap}\left(\theta_0\right)} U^{2}
\end{equation}

We see that at low voltages, the deviation from the Young
angle is proportional to $U^{2}$, just as in the Young-Lippmann formula.
However, the pre-factor is a function of $\theta_0$ and can take different
values than in eq~\ref{eq:Y-L_formula}, and even change its sign (see section II.D).
It is shown in section IV.D  under which conditions the pre-factor converges
to that of the Young-Lippmann formula for low voltages in typical EWOD
experimental setups.

%We show in section IV.D, where it is shown under which
%conditions the pre-factor converges to that of the Young-Lippmann formula for low
%voltages in typical EWOD experimental setups.

%%%%%%%%%%%%%%%%%%%%%%%%%%%%%%%%%%%%%%%%
\subsection{The High Voltage Limit}
%%%%%%%%%%%%%%%%%%%%%%%%%%%%%%%%%%%%%%%%
For $U\rightarrow\infty$, the electric energy becomes large relative to the capillary
energy and so the minimum of $F_{\rm tot}$ occurs at $\theta(U)=
\theta_{\rm sat}+\delta\theta$.
Expanding the condition $F'_{\rm tot}(\theta)=0$ around $\theta_{\rm sat}$, one has
\begin{equation}
F'_{\rm cap}\left(\theta_{\rm sat}\right)+ F''_{\rm cap}
\left(\theta_{\rm sat}\right)\delta\theta-\frac{1}{2}
 C''_{\rm tot}\left(\theta_{\rm sat}\right)U^{2}\delta\theta \simeq  0
\end{equation}
or
\begin{eqnarray}
\label{eq:highU}
\theta(U)& \simeq & \theta_{\rm sat}+ \frac{F'_{\rm cap}(\theta_{\rm sat})}
{\frac{1}{2}U^{2}C''_{\rm tot}(\theta_{\rm sat})-F''_{\rm cap}
(\theta_{\rm sat})}\nonumber\\
&\simeq& \theta_{\rm sat}+2\frac{F'_{\rm cap}
(\theta_{\rm sat})}{C''_{\rm tot}
(\theta_{\rm sat})}U^{-2}
\end{eqnarray}

Hence, saturation in $\theta$ is approached asymptotically, as $U^{-2}$,
in qualitative agreement with experiments~\cite{B2app_2005,Romi_2008}.

%%%%%%%%%%%%%%%%%%%%%%%%%%%%%%%%%%%%%%%%%%%%
\subsection{Reversed Electrowetting}
%%%%%%%%%%%%%%%%%%%%%%%%%%%%%%%%%%%%%%%%%%%%%%
An interesting conclusion can be drawn from
the discussion in section II.A. Recalling that in our model electrowetting
results from an interplay between capillary and electric energies (each with its
own minimum at $\theta_0$ and $\theta_{\mathrm{sat}}$, respectively),
as voltage is increased the electric energy gradually becomes
dominant and the contact angle is driven away from $\theta_0$
towards $\theta_{\mathrm{sat}}$ (figure~\ref{fig:saturation_sketch}). Since $\theta_0$
is determined only by the capillary parameters (as in the Young formula, eq~\ref{eq:Young})
and $\theta_{\mathrm{sat}}$ is determined solely by the electric
parameters, it is possible to envisage a system in which the
saturation angle $\theta_{\rm sat}$ is actually \textit{larger} than the Young angle,
$\theta_{\mathrm{sat}}>\theta_0$
rather than $\theta_{\mathrm{sat}}<\theta_0$ as in the
usual case. In such a setup, applying a voltage will cause {\it an increase} of the contact angle, in total contradiction with the Young-Lippmann
formula, eq~\ref{eq:Y-L_formula}. Hence, the model proposed here allows for the possible existence
of a new regime of electrowetting, which we refer to as {\it reversed
electrowetting}.

By examining the slopes of each energy term near the minimum of the other
(see figure~\ref{fig:min_energy_sketch}), it is possible to show that in the low- and high-voltage
limits, eqs~\ref{eq:lowU} and~\ref{eq:highU}, the pre-factors of both $U^{2}$ and $U^{-2}$ terms can
take either  positive or negative values depending on whether
$\theta_0>\theta_{\mathrm{sat}}$
or $\theta_0<\theta_{\mathrm{sat}}$; for the low-voltage
limit $F''_{\rm cap}\left(\theta_0\right)$ is positive
by definition, but $C'_{\rm tot}\left(\theta_0\right)$
is positive only if $\theta_0>\theta_{\mathrm{sat}}$ and negative
for $\theta_0<\theta_{\mathrm{sat}}$. Likewise, for the
high-voltage limit $C''_{\rm tot}\left(\theta_{\rm sat}\right)$
is negative by definition but $F'_{\rm cap}\left(\theta_{\rm sat}\right)$
is negative only if $\theta_0>\theta_{\mathrm{sat}}$ and positive
for $\theta_0<\theta_{\mathrm{sat}}$. Thus, we have shown how reversed electrowetting manifests itself in those limits.

%%%%%%%%%%%%%%%%%%%%%%%%%%%%%%%%%%%%%
\section{A Two Electrode Model of EWOD}
%%%%%%%%%%%%%%%%%%%%%%%%%%%%%%%%%%%%
Our goal in the remainder of this work is to elaborate on the physical conditions
that are involved in determining
a finite $\theta_{\rm sat}$ angle in {\it specific}  EWOD experimental setups.
However, we would like to stress that the proposed mechanism  is general and may be applied to
other realizations and experimental setups manifesting CAS.

%%%%%%%%%%%%%%%%%%%%%%%%%%%%%%%%%%%%%%%%
\subsection{System Setup \& Geometry}
\begin{figure}[h]
\begin{center}
\includegraphics[width=6cm]{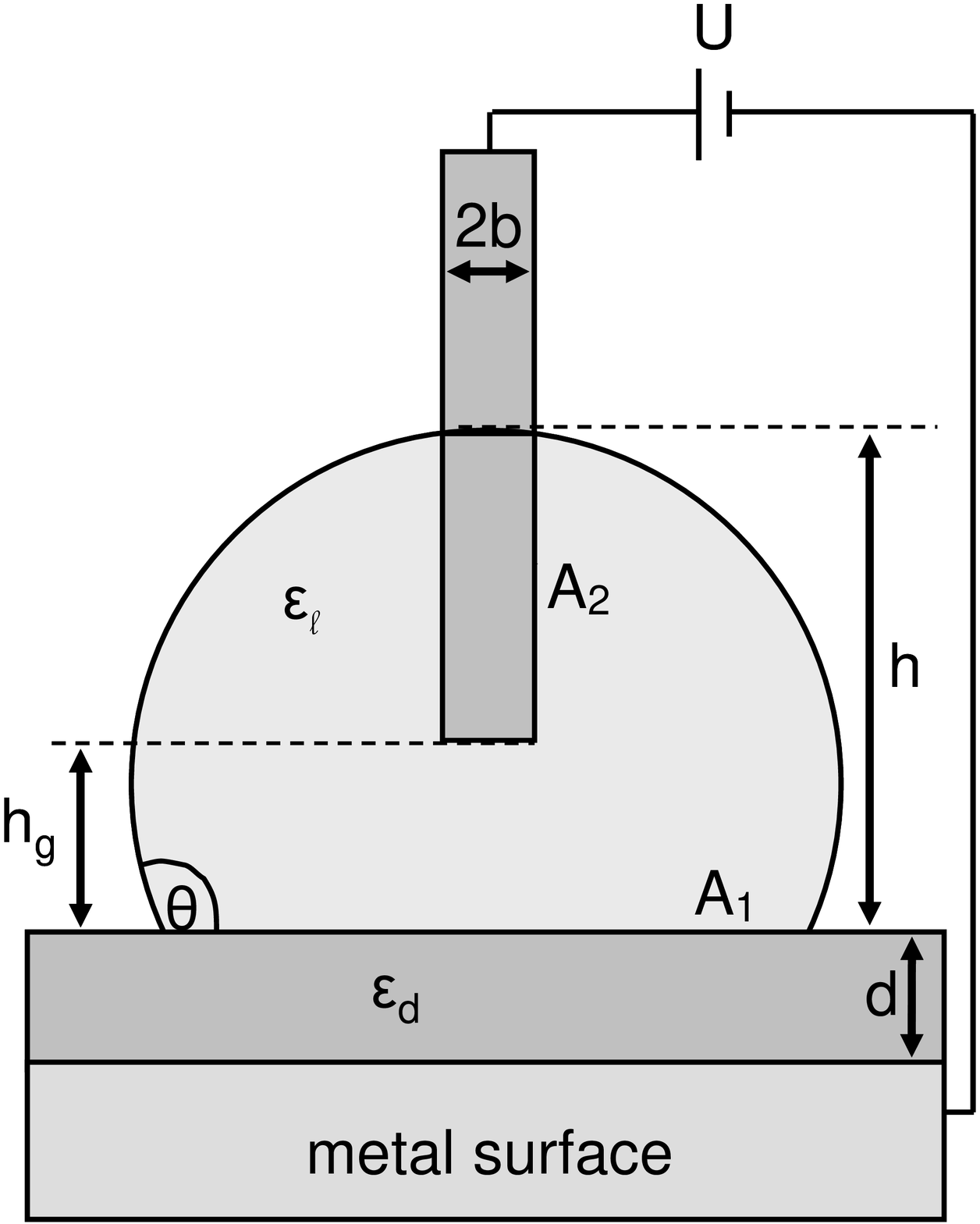}
\caption{\label{fig:Geometry}\textsf{A schematic EWOD
setup as used in our analysis. A liquid drop shaped as a spherical cap of volume $V$, height $h$
and dielectric constant $\varepsilon_l$   is placed atop  a flat metal electrode.
The metal electrode is covered with a dielectric coating of thickness $d$
and dielectric constant $\varepsilon_d$. A metal wire (used as a counter-electrode), modeled as
a thin cylinder
of radius $b$, is inserted into the drop from above. The gap between the two electrodes is $h_g$.
The area of the substrate electrode covered by the drop is $A_1$ and that of the counter-electrode is $A_2$.
The applied voltage is $U$ and the contact angle with the substrate is $\theta$.
}}
\end{center}
\end{figure}

%%%%%%%%%%%%%%%%%%%%%%%%%%%%%%%%%%%%%%%
A setup of an EWOD setup is presented
in figure~\ref{fig:Geometry}. The drop (dielectric constant $\varepsilon_l$)
is assumed to retain its spherical-cap shape, with height $h$ from the surface, total volume $V$
and contact angle $\theta$.  The metal electrode is coated with a dielectric layer of thickness
$d$ and dielectric constant $\varepsilon_d$. The top counter-electrode
is modeled as  a thin cylinder of radius $b$ and the gap between the two electrodes is $h_g$.
The two electrode areas covered by the liquid are $A_1$ and $A_2$, respectively.
For spherical-cap shaped drops $A_1$ and $A_2$ are related to the contact angle $\theta$ through the
fixed volume constraint:
\begin{eqnarray}
A_{1}&=&\pi a^{2} \nonumber\\
A_{2}&=&2\pi b(h-h_g) \nonumber\\
\tan\frac{\theta}{2}&=&\frac{h}{a} \nonumber\\
V&=&\frac{\pi h}{6}(3a^{2}+h^{2})
\label{eq:geometry}
\end{eqnarray}
where $a$ is the radius of the
covered portion of the substrate electrode.

It should be noted that $\theta$ can only take
values in the range $\theta_{\rm min}<\theta<\pi$.
%where $\theta_{\rm min}$
%is the angle at which the drop detaches from the counter-electrode.
The lower limit, $\theta_{\rm min}$, occurs when the drop height matches the gap between the lower tip
of the counter-electrode and the substrate, $h=h_g$ (see figure~\ref{fig:Geometry}), resulting in
\begin{equation}
\label{eq:theta_min}
\cot\frac{\theta_{\rm min}}{2}=\sqrt{\frac{2V}{\pi h_g^3}-\frac{1}{3}}
\end{equation}
We will show that
$\theta_{\rm sat}>\theta_{\rm min}$, hence
$\theta_{\rm min}$ is an inaccessible lower
bound of the contact angle. The upper limit $\theta=\pi$
is the de-wetting limit.

%%%%%%%%%%%%%%%%%%%%%%%%%%%%%%%%%%%%%%%%%%%%%%%%%%%%%%%%%%%%%%%%%%%%%%%%%%%%%%
%fig5
%%%%%%%%%%%%%%%%%%%%%%%%%%%%%%%%%%%%%%%%%%%%%%%%%%%%%%%%%%%%%%%%%%%%%%%%%%%%%%
%%%%%%%%%%%%%%%%%%%%%%%%%%%%%%%%%%%%%%%%%%%%%%%%%%%%%%%%%%%%%%%%%%%%%%%%%%%%%%

%%%%%%%%%%%%%%%%%%%%%%%%%%%%%%%%%%%%%%%%
\subsection{The AC Free-Energy}
%%%%%%%%%%%%%%%%%%%%%%%%%%%%%%%%%%%%%%%%
The free energy, eq~\ref{eq:generalized_freeE},
depends on the total capacitance $C_{\rm tot}$, which includes
all relevant contributions.
Unlike the traditional
Young-Lippmann treatment in which only the capacitance of the liquid/substrate interface is taken
into account, we  consider explicitly the existence of an additional double-layer,
residing at the interface between the liquid drop and the counter-electrode.

Experimental setups and applications usually employ AC circuits
to produce an electrowetting effect. Under those circumstances,
double-layers are transient: with each AC half-cycle  a double-layer
of opposite polarity is formed at each electrode/liquid interface and subsequently dissolved away.
In addition to relaxation by
reversal of polarity, other mechanisms of relaxation can act at the
counter-electrode/liquid interface such as electrochemical Faradic
processes. The dynamical processes are, therefore, governed by two  intrinsic time scales (beside the AC frequency):
%
%%%%%%%%%%%%%%%%%%%%%%
\begin{enumerate}
%%%%%%%%%%%%%%%%%%%%%%
\item
The double-layer {\it build-up time}, $\tau_{b}$,  which can be estimated to be
\begin{equation}
\tau_{b}\simeq \frac{\lambda_{D} L}{D}
\label{eq:tau}
\end{equation}
where $\lambda_{D}=\sqrt{\varepsilon_0\varepsilon_l k_BT/2 e^2 c_{\rm salt}}$
is the Debye length, $k_BT$ the thermal energy, $c_{\rm salt}$ the salt concentration, $D$ is
the diffusion constant and $L$ is a typical system size~\cite{Bazant_2004}.

\item
The double-layer {\it relaxation time}, $\tau_{r}$,  which can similarly be expressed in terms
of system parameters through the RC circuit relaxation formula
\begin{equation}
\tau_{r}=R_2C_2=\rho\frac{C_2}{A_2}
\label{eq:Tr}
\end{equation}
where $\rho=R_2 A_2$ is defined as  the zero-current (Faradic) resistivity to charge transfer by
electrochemical processes and $A$ is the contact area.
\end{enumerate}

In order to discuss the period-averaged properties of the system, we employ a standard
AC circuit analysis. As shown in figure~\ref{fig:AC_equiv_circ}(b),
we model the two liquid/electrode interfaces as two capacitors with capacitances ${C_{1},\, C_{2}}$,
defined in a similar fashion as in eq~\ref{eq:Cld}:
\begin{eqnarray}
\label{eq:C12}
C_{1}&=&C_{\rm ld}\simeq \frac{\varepsilon_{0}\varepsilon_{d}A_{1}}{d} \nonumber\\
C_{2}&\simeq &\frac{\varepsilon_{0}\varepsilon_{l}A_{2}}{\lambda_{D}}
\end{eqnarray}
Note that  the main contribution to $C_1$ comes from the coated dielectric layer of thickness $d$
($d/\varepsilon_l\gg \lambda_D/\varepsilon_l$),
while for $C_2$, the only contribution
comes from the double layer of thickness $\lambda_D$ (because the counter-electrode is {\it not} coated).
The cylindrical geometry of the counter-electrode is not considered  because $b\gg \lambda_D$.

The two capacitors are charged and discharged through a resistor $R_{1}$ that represents
the bulk of the liquid drop. The  relaxation of the double-layer at
the counter-electrode is modeled by an extra discharge circuit with a resistor
$R_{2}$, while the capacitor $C_{1}$ does not have a discharge circuit
since charge transfer at the substrate electrode is prevented by
its dielectric coating. The appropriate resistance values can
be inferred from the build-up ($\tau_b$) and relaxation ($\tau_r$)
times, eqs~\ref{eq:tau} and \ref{eq:Tr}, again through the RC circuit relaxation formula:
\begin{eqnarray}
  R_{1}&=&\tau_{b}(C_{1}^{-1}+C_{2}^{-1})\nonumber\\
R_{2}&=& \frac{\tau_{r}}{C_{2}}
\label{eq:time_scales}
\end{eqnarray}

Drawing on the AC circuit analogy, the period-averaged free energy is:
\begin{equation}
F_{\rm tot}(\theta,U,\omega)=F_{\rm cap}+F_{\rm el}
=F_{\rm cap}(\theta)- \frac{1}{2}\frac{1}{\omega\left|Z_{\rm tot}(\theta,\omega)\right|}U^{2}\label{eq:ACmodelE}
\end{equation}
where $U$ is understood to be the rms value and $Z_{\rm tot}$ is the total impedance of the circuit, [figure~\ref{fig:AC_equiv_circ}(b)],
which can be represented schematically as
\begin{equation}
Z_{\rm tot}=Z_{C_1}\oplus Z_{R_1}\oplus (Z_{C_2}||Z_{R_2})
\end{equation}
It is straight forward to show that the squared magnitude of the total
impedance is
\begin{eqnarray}
\label{eq:Z_AC_model}
\left|Z_{\rm tot}\right|^{2}&=&\frac{1}{C_{1}^{2}}
\left(\tau_{b}^{2}+\omega^{-2}\right)+\frac{2}{C_{1}C_{2}}
\left(\tau_{b}^{2}+\frac{\tau_{r}(\tau_{r}+\tau_{b})}{1+ \omega^{2}\tau_{r}^{2}}\right)\nonumber\\
&+&\frac{1}{C_{2}^{2}}\left(\tau_{b}^{2}+\frac{\tau_{r}(\tau_{r}+2\tau_{b})}{1+ \omega^{2}\tau_{r}^{2}}\right)
\end{eqnarray}

Since $C_{1}$ is proportional to $A_{1}$, which  vanishes at $\theta\rightarrow\pi$
and $C_{2}$ is proportional to $A_{2}$, which  vanishes at $\theta\rightarrow\theta_{\mathrm{min}}$,
$\left|Z_{\mathrm{tot}}\right|^{2}$ diverges both at $\theta\rightarrow\theta_{\mathrm{min}}$
and $\theta\rightarrow\pi$.
Therefore, it must have a minimum at some intermediate value: $\theta_{\mathrm{min}}<\theta_{\mathrm{sat}} <\pi$.
Hence, our model of electrowetting presented in section II.A
is indeed applicable to typical EWOD setups.

Substituting eqs~\ref{eq:geometry} and~\ref{eq:C12} into eqs~\ref{eq:ACmodelE} and~\ref{eq:Z_AC_model} yields an expression for
$C_1(\theta)$ and $C_2(\theta)$  and, consequently,  for $F_{\rm el}$ as a function of $\theta$.
Its minimization can be done numerically (section~IV.A) and yields the equilibrium contact angle
(for given applied voltage and frequency). In some limits (sections~IV.C and IV.E)  analytical approximations can be derived as well.

%%%%%%%%%%%%%%%%%%%%%%%%%%%%%%%%
\section{Results and Discussion}
%%%%%%%%%%%%%%%%%%%%%%%%%%%%%%%%%

%%%%%%%%%%%%%%%%%%%%%%%%%%%%%%%%%%%%%%%%%%%%%%%
\subsection{The Electrowetting Curve, $\theta(U)$}
%%%%%%%%%%%%%%%%%%%%%%%%%%%%%%%%%%%%%%%%%%%%%%%

\begin{figure}[h]
\begin{center}
\includegraphics[width=8cm]{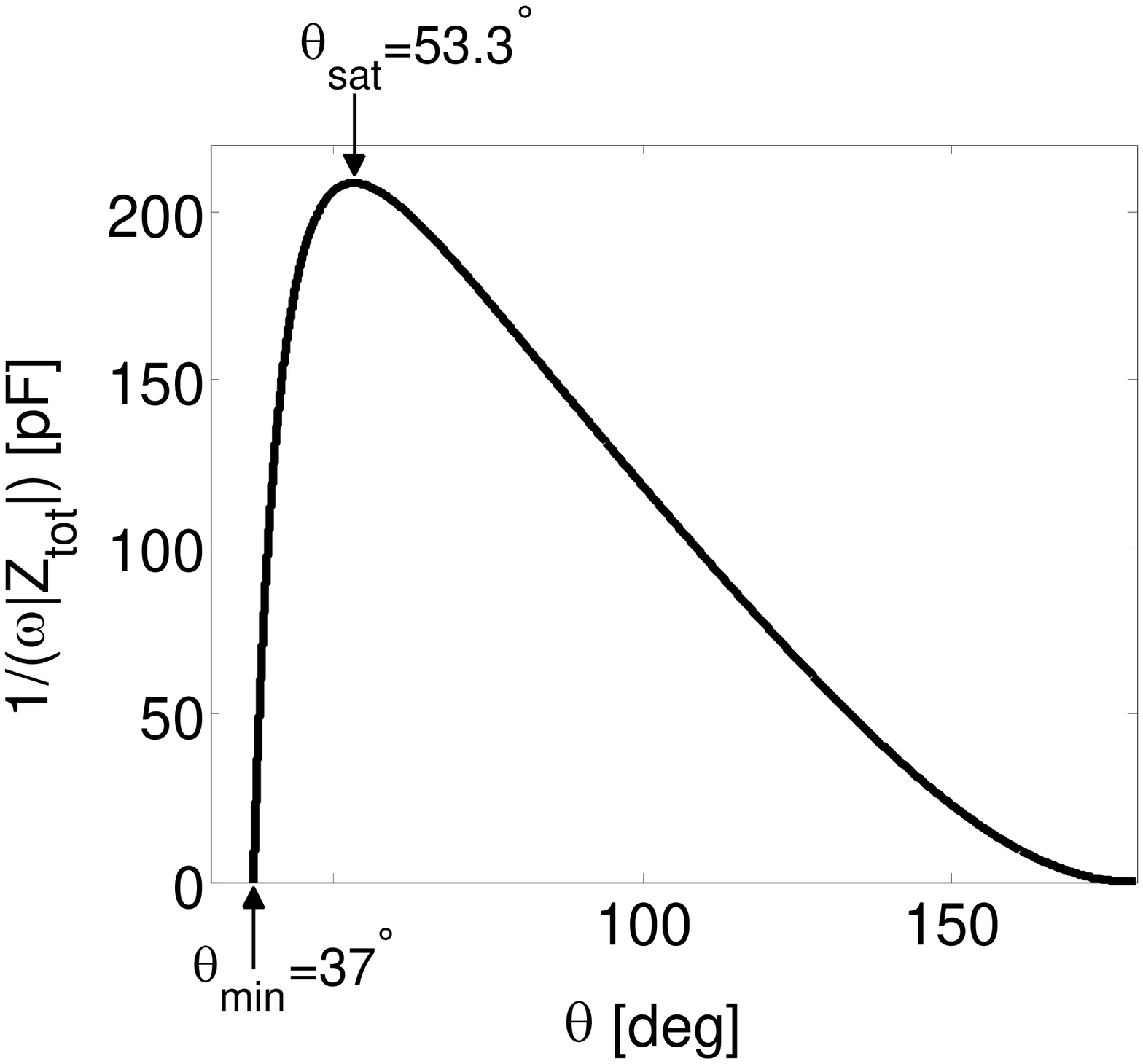}
\caption{\label{fig:AC_el}\textsf{Total reactance $1/(\omega |Z_{\rm tot}|)$
(in nanoFarad) as a function of the contact angle. Parameters for the
EWOD system are $\tau_{b}=1.34$\,ms, $\tau_{r}=0.53$\,s
 and $f=\omega/2\pi=1$\,kHz, and all other parameters are chosen as in table~\ref{tab:system}.
The reactance has a maximum for a finite value of $\theta_{\rm sat}=53.3^{\circ}$, at which
the system exhibits CAS.
The minimal contact angle is $\theta_{\rm min}=37^{\circ}$}, eq~\ref{eq:theta_min}.}
\end{center}
\end{figure}

In order to demonstrate quantitatively the model validity, we performed numerical
calculations for parameter values that are in accord with
some typical experimental setups. In figure~\ref{fig:AC_el} we present the reactance
$1/\omega Z$ (appearing in eq~\ref{eq:ACmodelE}) computed for parameter values as
in table~\ref{tab:system} and with an AC frequency  $f=\omega/2\pi=1$\,kHz. The build-up time
was calculated using eq~\ref{eq:tau} to be $\tau_{b}=1.34$\,ms, while the relaxation time was
calculated using eq~\ref{eq:Tr} with~\cite{ft7} $\rho=1\,\Omega\cdot m^2$, yielding~\cite{ft6}
$\tau_{r}=0.53$\,s. For the chosen values of parameters,
the ratio of capacitances for zero voltages ($\theta(U=0)=\theta_0$)
is about, $C_2/C_1\simeq 25$. In the figure a maximum at a finite angle $\theta_{\rm sat}=53.3^{\circ}$ is
clearly seen. Notably, this saturation angle is much larger than the minimal possible angle in this setup,
$\theta_{\rm min}=37^{\circ}$ (see eq~\ref{eq:theta_min}). As a consequence
CAS is obtained for finite values of $A_2$, much before the limit $A_2\to 0$ characteristic to $\theta_{\rm min}$.

%%%%%%%%%%%%%%%%%%%%%%%%%%%%%%%%%%%%%%%%%%%%%%%%%%%%%%%%%%%%%%%%%%%%%%%%%%%%%%
%fig6.eps
%%%%%%%%%%%%%%%%%%%%%%%%%%%%%%%%%%%%%%%%%%%%%%%%%%%%%%%%%%%%%%%%%%%%%%%%%%%%%%
%%%%%%%%%%%%%%%%%%%%%%%%%%%%%%%%%%%%%%%%%%%%%%%%%%%%%%%%%%%%%%%%%%%%%%%%%%%%%%

%%%%%%%%%%%%%%%%%%%%%%%%%%%%%%%%%%%%%%%%%%%%%%%%%%%%%%%%%%%%%%%%%%%%%%%%%%%%%%
%Table 1
%%%%%%%%%%%%%%%%%%%%%%%%%%%%%%%%%%%%%%%%%%%%%%%%%%%%%%%%%%%%%%%%%%%%%%%%%%%%%%%
\begin{table}[tp]
\begin{centering}
\begin{tabular}{|l|c|l|}
\hline
~Parameter  & ~symbol  & ~value\tabularnewline
\hline
\hline
~dielectric constant of liquid  & $\varepsilon_{l}$  & ~$80$\tabularnewline
\hline
~Debye length in liquid  & $\lambda_{D}$  & ~$1.34\textrm{nm}$\tabularnewline
\hline
~volume  & $V$  & ~$5\mu\textrm{L}$\tabularnewline
\hline
~width of dielectric layer  & $d$  & ~$0.1\mu\textrm{m}$\tabularnewline
\hline
~dielectric constant of dielectric layer  & $\varepsilon_{d}$  & ~$2.67$\tabularnewline
\hline
~liquid/air surface tension  & $\gamma_{\rm la}$  & ~$72.8\textrm{mN/m}$\tabularnewline
\hline
~dielectric/air surface tension  & $\gamma_{\rm sa}$  & ~$12.7\textrm{mN/m}$\tabularnewline
\hline
~liquid/dielectric interfacial tension  & $\gamma_{\rm sl}$  & ~$47\textrm{mN/m}$\tabularnewline
\hline
~gap between counter-electrode and substrate  & $h_g$  & ~$0.7\textrm{mm}$\tabularnewline
\hline
~radius of counter-electrode  & $b$  & ~$12.5\mu\textrm{m}$\tabularnewline
\hline
\multicolumn{1}{c}{} & \multicolumn{1}{c}{} & \multicolumn{1}{c}{}\tabularnewline
\end{tabular}
\par\end{centering}
\caption{\label{tab:system}\textsf{Parameter values of a typical electrowetting setup. The liquid drop
contains an aqueous ionic solution and is placed on top of a Miyaline-C/Teflon substrate.}}
\end{table}
%%%%%%%%%%%%%%%%%%%%%%%%%%%%%%%%%%%%%%%%%%%%%%%%%%%%%%%%%%%%%%%%%%%%%%%%%%%%%%

\begin{figure}[h]
\begin{center}
\includegraphics[width=8cm]{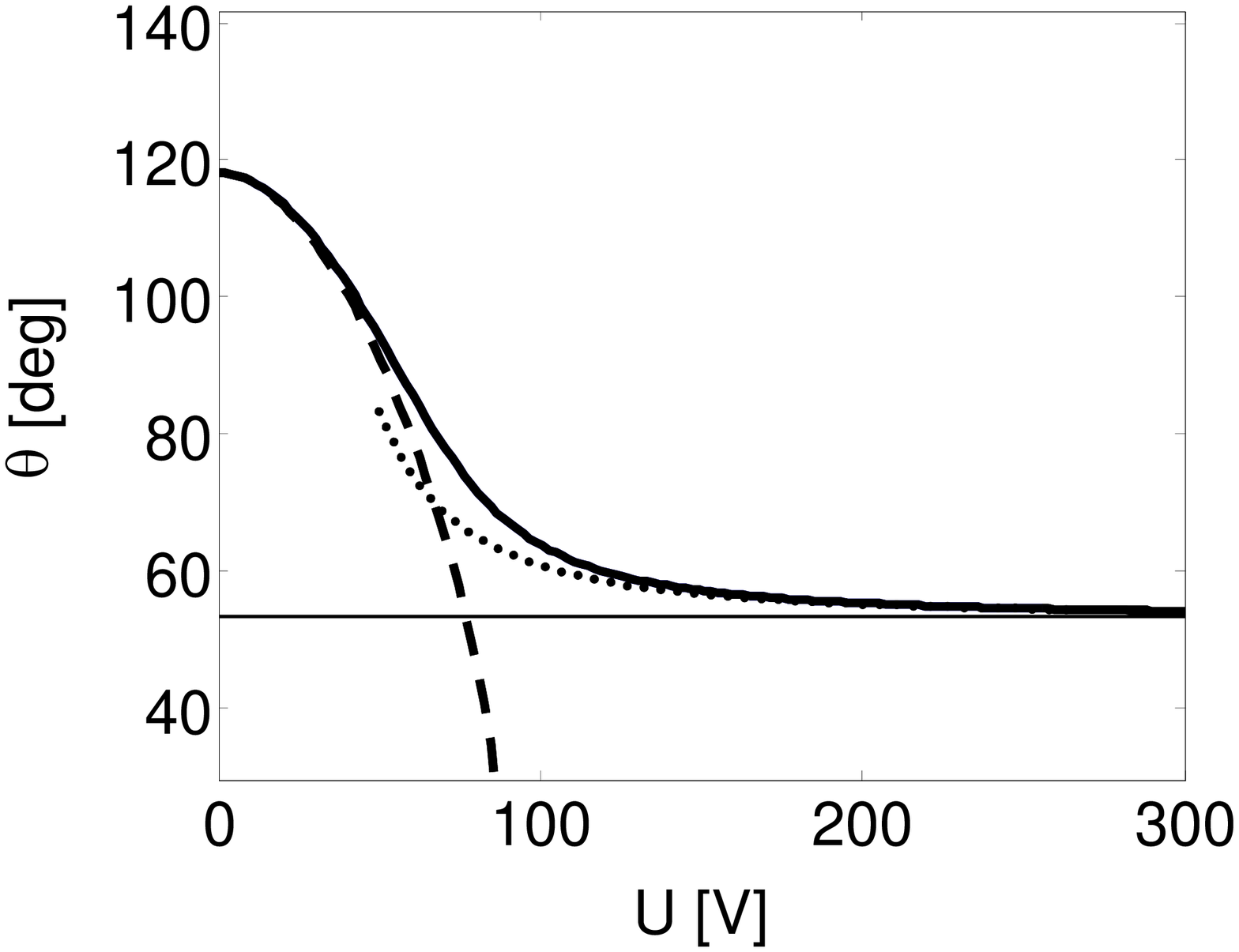}
\caption{\label{fig:ACew}\textsf{Calculated contact angle as a function of applied
voltage, $U$ (full line), as well as a plot of the  Young-Lippmann formula using $g^{\rm eff}$
(dashed line, see text). Parameter values of the EWOD system are taken from table~\ref{tab:system}
and $\tau_{b}=1.34$\,ms, $\tau_{r}=0.53$\,s and
$f=\omega/2\pi=1$\,kHz. Many of the features of CAS,   shown schematically in figure~\ref{fig:exp_sketch},
are reproduced.
At $U=0$, the contact angle is $\theta_{0}=118^{\circ}$.  For small $U$, there is an initial
compliance with the rescaled Young-Lippmann formula, eq~\ref{eq:Y-L_formula}, $\theta(U)\approx \theta_{\rm YL}(U)$,
followed by a cross-over occurring
at $U^*=76.9$\,V, calculated from eq~\ref{eq:U_cross-over}.  At larger $U$,
the contact angle tends asymptotically to a saturation angle $\theta_{\rm sat}=53.3^{\circ}$.
The asymptotic $\theta(U)-\theta_{\rm sat}\sim U^{-2}$ is plotted (dotted line) following eq \ref{eq:highU} and approximates rather well $\theta(U)$ for voltages larger than 120\,V.
An operational definition of the saturation voltage $U_{\rm sat}$ (see text) yields $U_{\rm sat}\simeq 252.1$\,V.
Note that $\theta_{\rm sat}$ is conceivably larger than
the minimal possible angle $\theta_{\rm min}=37^{\circ}$, eq~\ref{eq:theta_min}.}}
\end{center}
\end{figure}

Figure~\ref{fig:ACew} presents the calculated electrowetting curve $\theta(U)$
for the same system, where $\theta(U)$ is calculated by minimizing $F_{\rm tot}$
from eq~\ref{eq:ACmodelE}, together with a plot of the Young-Lippmann formula (dashes), where an effective $g^{\rm eff}$ pre-factor
is used to
fit the full calculation. This is similar to what is done in many experimental works where the $g$ value is fitted from the low $U$ dependence,
and not by using explicitly eq~\ref{eq:Y-L_formula}. We use a specific $g^{\rm eff}=(1+\omega^2\tau_b^2)^{-1/2}$ as
derived in section~IV.D.
The figure shows that several common experimental
features are reproduced (as compared with the schematic figure~\ref{fig:exp_sketch}); an initial
compliance with the scaled Young-Lippmann formula at low voltages
is followed by a cross-over at intermediate voltages to a different regime. Using eq~\ref{eq:U_cross-over} (with $g^{\rm eff}$),
the cross-over voltage is evaluated to be $U^*\simeq 76.9$\,V.
At $U>U^*$ an asymptotic convergence  of the contact angle towards a saturation value is seen,
$\theta(U)\to \theta_{\rm sat}=53.3^{\circ}$. This is further demonstrated in figure~\ref{fig:ACew}, where the
asymptotic $\theta(U)-\theta_{\rm sat}\sim U^{-2}$ is plotted (dotted line) following eq \ref{eq:highU}. The asymptotic behavior approximates rather well $\theta(U)$ for voltages larger than 120\,V.

It is appropriate to define another voltage, $U_{\rm sat}$, characterizing the saturation range of the potential.
An operational  definition that we employ is that at $U_{\rm sat}$,
the calculated $\theta(U_{\rm sat})$ deviates from $\theta_{\rm sat}$ by 2\%. With this definition,
we obtain $U_{{\rm sat}}\simeq 252.1$\,V.
The electrowetting curve presented in figure~\ref{fig:ACew} agrees qualitatively with experimental observations \cite{FreqEW_2007,B2app_2005,Polarity_2006}, which show the effect of contact angle saturation. Unfortunately, because
the parameter values needed for quantitatively comparison with experiments lack at present,  we used instead reasonable estimations.
%

%%%%%%%%%%%%%%%%%%%%%%%%%%%%%%%%%%%%%%%%%%%%%%%%%%%%%%%%%%%%%%%%%%%%%%%%%%%%%%
%fig7.eps
%%%%%%%%%%%%%%%%%%%%%%%%%%%%%%%%%%%%%%%%%%%%%%%%%%%%%%%%%%%%%%%%%%%%%%%%%%%%%%
%%%%%%%%%%%%%%%%%%%%%%%%%%%%%%%%%%%%%%%%%%%%%%%%%%%%%%%%%%%%%%%%%%%%%%%%%%%%%%

%%%%%%%%%%%%%%%%%%%%%%%%%%%%%%%%%%%%%%%%%%%%%%%%%%%
\subsection{The Reversed Electrowetting Curve, $\theta(U)$}
%%%%%%%%%%%%%%%%%%%%%%%%%%%%%%%%%%%%%%%%%%%%%%%%%%%%%

We now illustrate how reversed electrowetting $\theta_{\rm sat}<\theta_0$, which
is a natural outcome of our model, can be seen in the laboratory. Let us consider a system similar to the one presented in the previous section with the two following changes (see table~\ref{tab:reversed_EW}): the interfacial tensions are chosen such that $\theta_0=60^{\circ}$, and we now model a non-polar liquid with a dielectric constant $\epsilon_{l}=2$, which yields a build-up time of $\tau_b=1.34$\,ms and a relaxation time of $\tau_r=13.2$\,ms. The AC frequency is $f=\omega/2\pi=1$\,kHz as before.

\begin{figure}[h]
\begin{center}
\includegraphics[width=8cm]{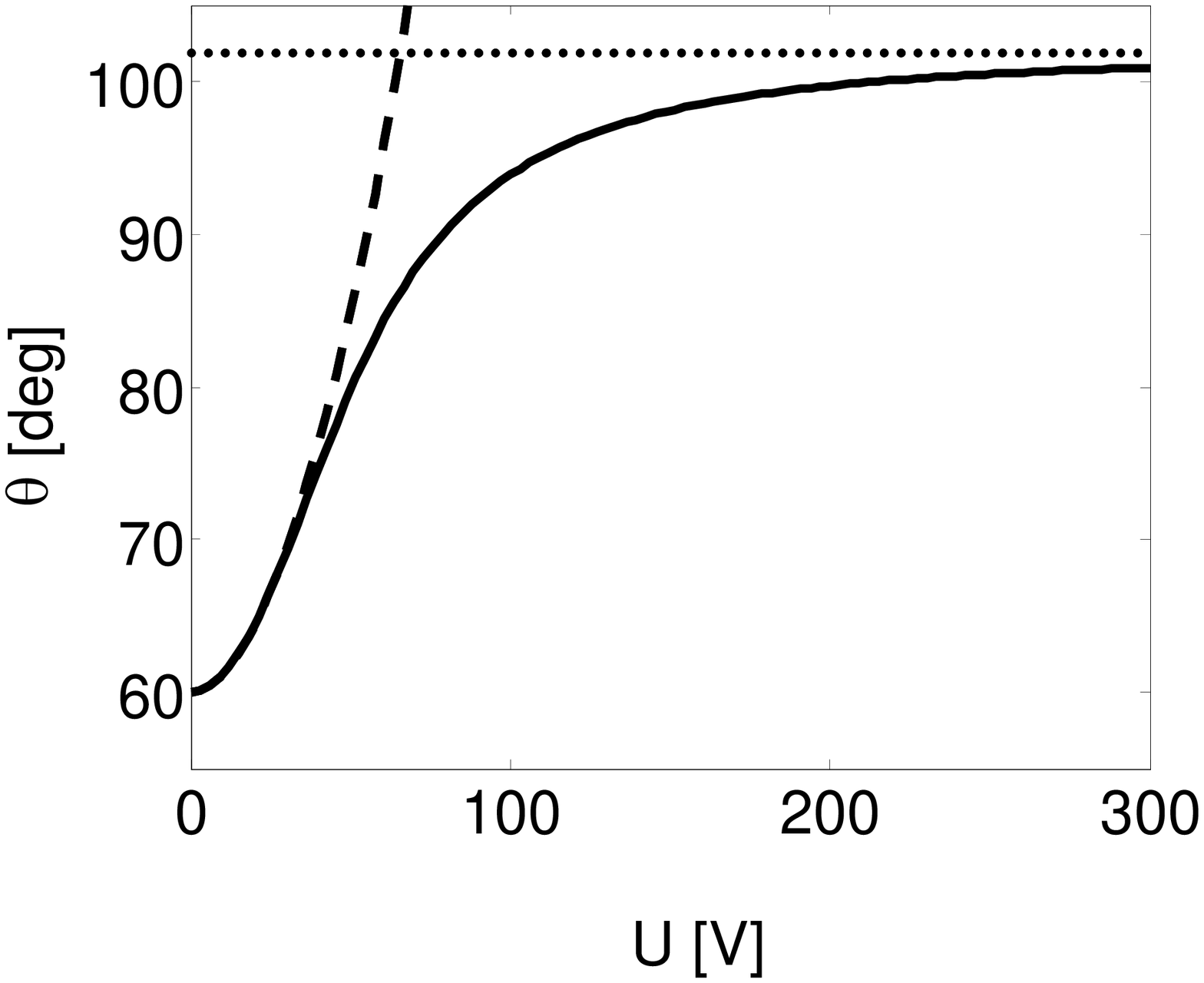}
\caption{\label{fig:reversed_EW}\textsf{Calculated contact angle as a function of applied
voltage, $U$, (solid line) for reversed electrowetting together with a manually scaled ($g^{\rm eff}<0$)
Young-Lippmann formula (dashed line). Parameter values of a EWOD system are taken from table~\ref{tab:reversed_EW}
and $\tau_{b}=1.34$\,ms, $\tau_{r}=13.2$\,ms and
$f=\omega/2\pi=1$\,kHz.
At $U=0$, the contact angle is $\theta_{0}=60^{\circ}$.  Even for small $U$, there is a deviation from
the naive Young-Lippmann formula, eq~\ref{eq:Y-L_formula}, because $\theta(U)-\theta_0\sim U^2$, with a positive pre-factor.
The $U^2$ rise is followed by a cross-over occurring
at $U^*=65.5$\,V.  At larger $U$,
the contact angle tends asymptotically to a saturation angle $\theta_{\rm sat}=101.9^{\circ}$.
Using the same definition of the saturation voltage $U_{\rm sat}$ as in figure~\ref{fig:ACew}
yields $U_{\rm sat}\simeq 208.5$\,V.}}
\end{center}
\end{figure}

Figure~\ref{fig:reversed_EW} presents the  calculated electrowetting curve $\theta(U)$. The plot features an initial
compliance with the negatively rescaled Young-Lippmann formula (at low voltages such that the contact angle {\it increases}
with the applied voltage. This is followed by a cross-over at intermediate voltages towards saturation. Using eq~\ref{eq:U_cross-over} (with $g^{\rm eff}$),
the cross-over voltage is evaluated to be $U^*\simeq 65.5$\,V.
For $U>U^*$ an asymptotic convergence  of the contact angle towards a saturation value is seen,
$\theta(U)\to \theta_{\rm sat}=101.9^{\circ}$. Using the same definition as in section IV.A,
the saturation voltage is found to be $U_{{\rm sat}}\simeq 208.5$\,V.

Since our reversed electrowetting predictions (figure~\ref{fig:ACew}) are rather for specific parameter values, it will be of benefit to check their validity with experiments conducted on similar electrowetting setups.

%%%%%%%%%%%%%%%%%%%%%%%%%%%%%%%%%%%%%%%%%%%%%%%%%%%%%%%%%%%%%%%%%%%%%%%%%%%%%%
%Table 2 REVERSED EW
%%%%%%%%%%%%%%%%%%%%%%%%%%%%%%%%%%%%%%%%%%%%%%%%%%%%%%%%%%%%%%%%%%%%%%%%%%%%%%
\begin{table}[tp]
\begin{centering}
\begin{tabular}{|c|l|}
\cline{1-2}
~Parameter  & ~Value  \tabularnewline
\cline{1-2}
~$\varepsilon_{l}$  & ~$2$   \tabularnewline
\cline{1-2}
~$\lambda_{D}$  & ~$1.34\textrm{ nm}$   \tabularnewline
\cline{1-2}
~$V$  & ~$5\textrm{ \ensuremath{\mu}L}$ \tabularnewline
\cline{1-2}
~$d$  & ~$0.1\textrm{ \ensuremath{\mu}m}$  \tabularnewline
\cline{1-2}
~$\varepsilon_{d}$  & ~$2.67$  \tabularnewline
\cline{1-2}
~$\gamma_{\rm la}$  & ~$20\textrm{ mN/m}$\tabularnewline
\cline{1-2}
~$\gamma_{\rm sa}$  & ~$15\textrm{ mN/m}$\tabularnewline
\cline{1-2}
~$\gamma_{\rm sl}$  & ~$5\textrm{ mN/m}$\tabularnewline
\cline{1-2}
~$h_g$  & ~$0.7\textrm{ mm}$\tabularnewline
\cline{1-2}
~$b$ & ~$12.5\textrm{ \ensuremath{\mu}m}$\tabularnewline
\cline{1-2}
\multicolumn{1}{c}{} & \multicolumn{1}{c}{}
\tabularnewline
\end{tabular}
\par\end{centering}
\caption{\label{tab:reversed_EW}\textsf{Parameter values of a hypothetical system that exhibits reversed electrowetting.}}
\end{table}
%%%%%%%%%%%%%%%%%%%%%%%%%%%%%%%%%%%%%%%%%%%%%%%%%%%%%%%%%%%%%%%%%%%%%%%%%%%%%%

%%%%%%%%%%%%%%%%%%%%%%%%%%%%%%%%%%%%%%%%%%%%%%%%%%%%%%%%%%%%%%%%%%%%%%%%%%%%%
%fig8 (Reversed EW)
%%%%%%%%%%%%%%%%%%%%%%%%%%%%%%%%%%%%%%%%%%%%%%%%%%%%%%%%%%%%%%%%%%%%%%%%%%%%%%
%%%%%%%%%%%%%%%%%%%%%%%%%%%%%%%%%%%%%%%%%%%%%%%%%%%%%%%%%%%%%%%%%%%%%%%%%%%%%%

%%%%%%%%%%%%%%%%%%%%%%%%%%%%%%%%%%%%%%%%%%%%%%%%%%%
\subsection{The Frequency Dependence of Electrowetting}
%%%%%%%%%%%%%%%%%%%%%%%%%%%%%%%%%%%%%%%%%%%%%%%%%%%%%

\begin{figure}[h]
\begin{center}
\includegraphics[width=8cm]{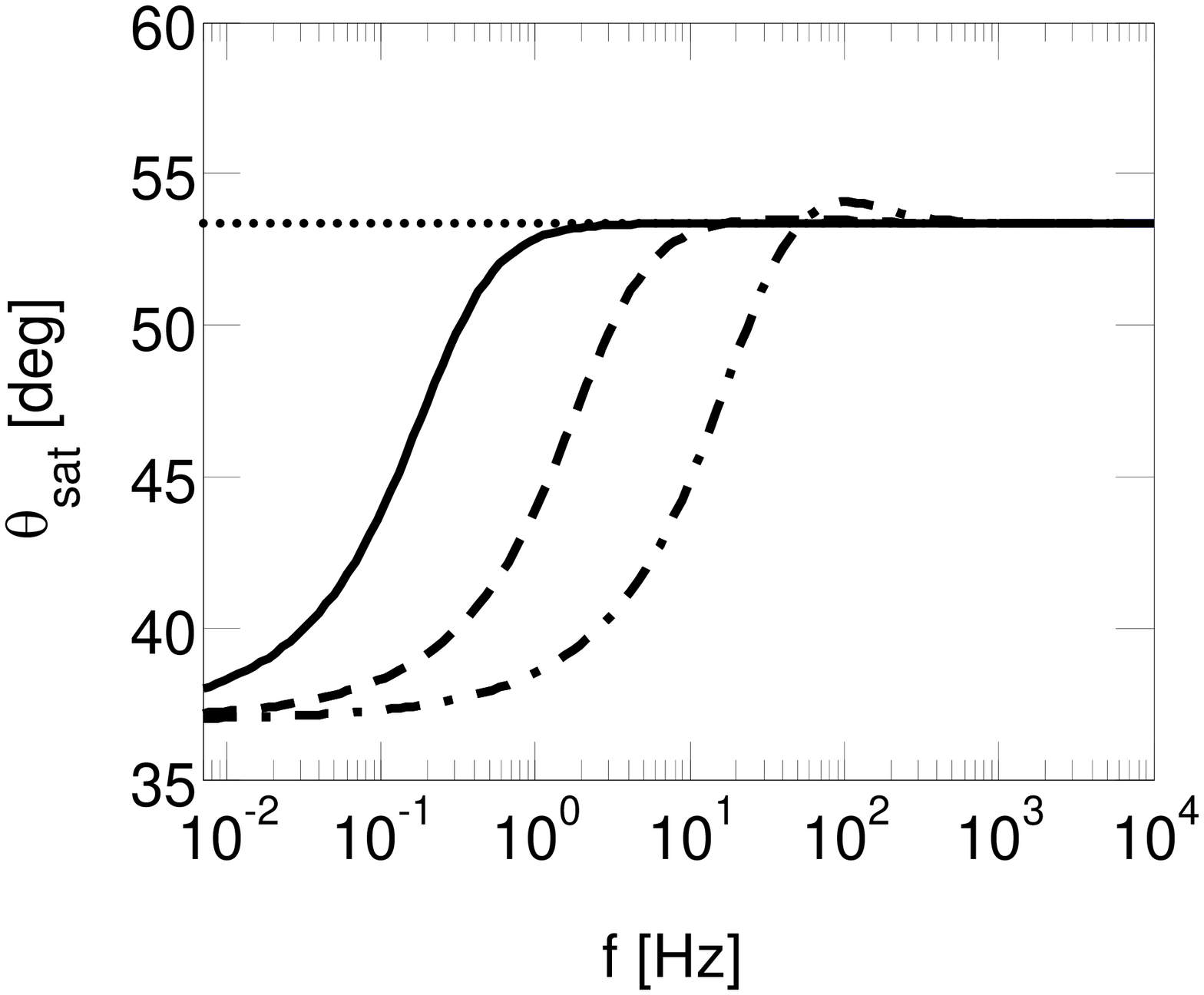}
\caption{\label{fig:theta_sat_f} \textsf{Saturation angle, $\theta_{\rm sat}$,
as function of the AC frequency $f$, for the EWOD system with parameter values as in table~\ref{tab:system},
$\tau_{b}=1.34$\,ms and for several  $\tau_{r}$ values: $\tau_r=0.53$\,s (solid line), 0.053\,s (dashed line),
5.3\,ms (dashed-dotted line).
In the range  $f\ge 1$\,{kHz}, the AC saturation angle, which
takes into account relaxation of the double-layer by electrochemical processes,
matches the $\tau_r\to\infty$ (dotted line) of $\theta_{\rm sat}^\infty=53.3^{\circ}$
even for the smallest $\tau_r=5.3$\,ms (fastest Faradic relaxation). }}
\end{center}
\end{figure}

In order to explore the effect of the frequency of the AC voltage within our model, the dependence
of the saturation angle $\theta_{\rm sat}$ on frequency
was calculated numerically by minimizing eq~\ref{eq:ACmodelE}  and plotted in  figure~\ref{fig:theta_sat_f} for several
values of $\tau_{r}=0.53$\,s, 0.053\,s, and 5.3\,ms.
It can be seen that for this specific choice of parameters, the AC saturation
angle reaches a constant value for the entire high frequency range
down to $f\simeq 1$\,kHz, even
for the smallest of the chosen relaxation times ($\tau_{r}=5.3$\,ms). Moreover,
the larger $\tau_{r}$ is, the wider is the range for which the saturation angle
is constant. This can be explained by taking into account that whenever $\tau_r\gg 1/f$, the counter-electrode double-layer can hardly relax.

%%%%%%%%%%%%%%%%%%%%%%%%%%%%%%%%%%%%%%%%%%%%%%%%%%%%%%%%%%%%%%%%%%%%%%%%%%%%%%
%fig9.eps
%%%%%%%%%%%%%%%%%%%%%%%%%%%%%%%%%%%%%%%%%%%%%%%%%%%%%%%%%%%%%%%%%%%%%%%%%%%%%%
%%%%%%%%%%%%%%%%%%%%%%%%%%%%%%%%%%%%%%%%%%%%%%%%%%%%%%%%%%%%%%%%%%%%%%%%%%%%%%

\begin{figure}[h]
\begin{center}
\includegraphics[width=8cm]{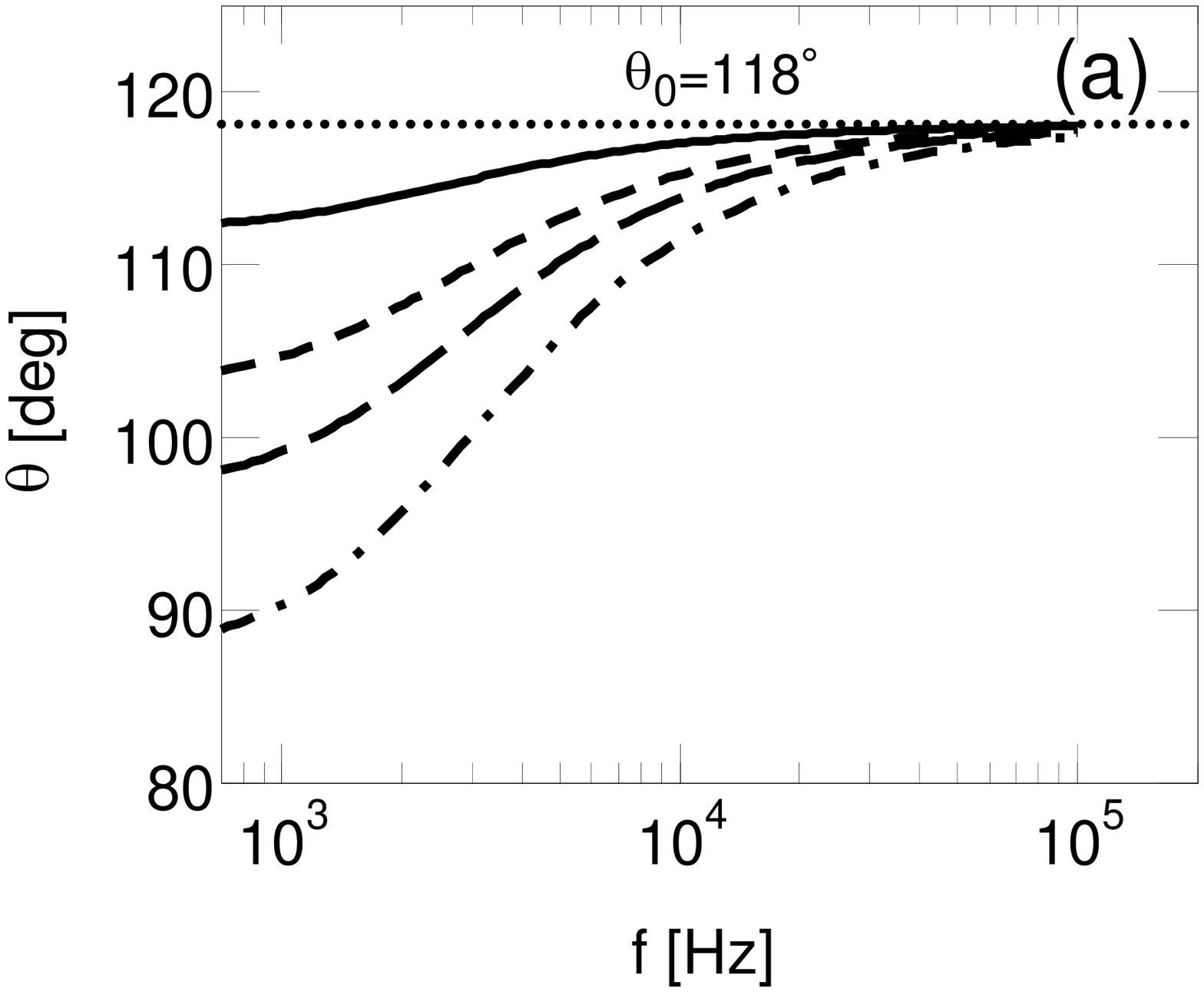}
\includegraphics[width=8cm]{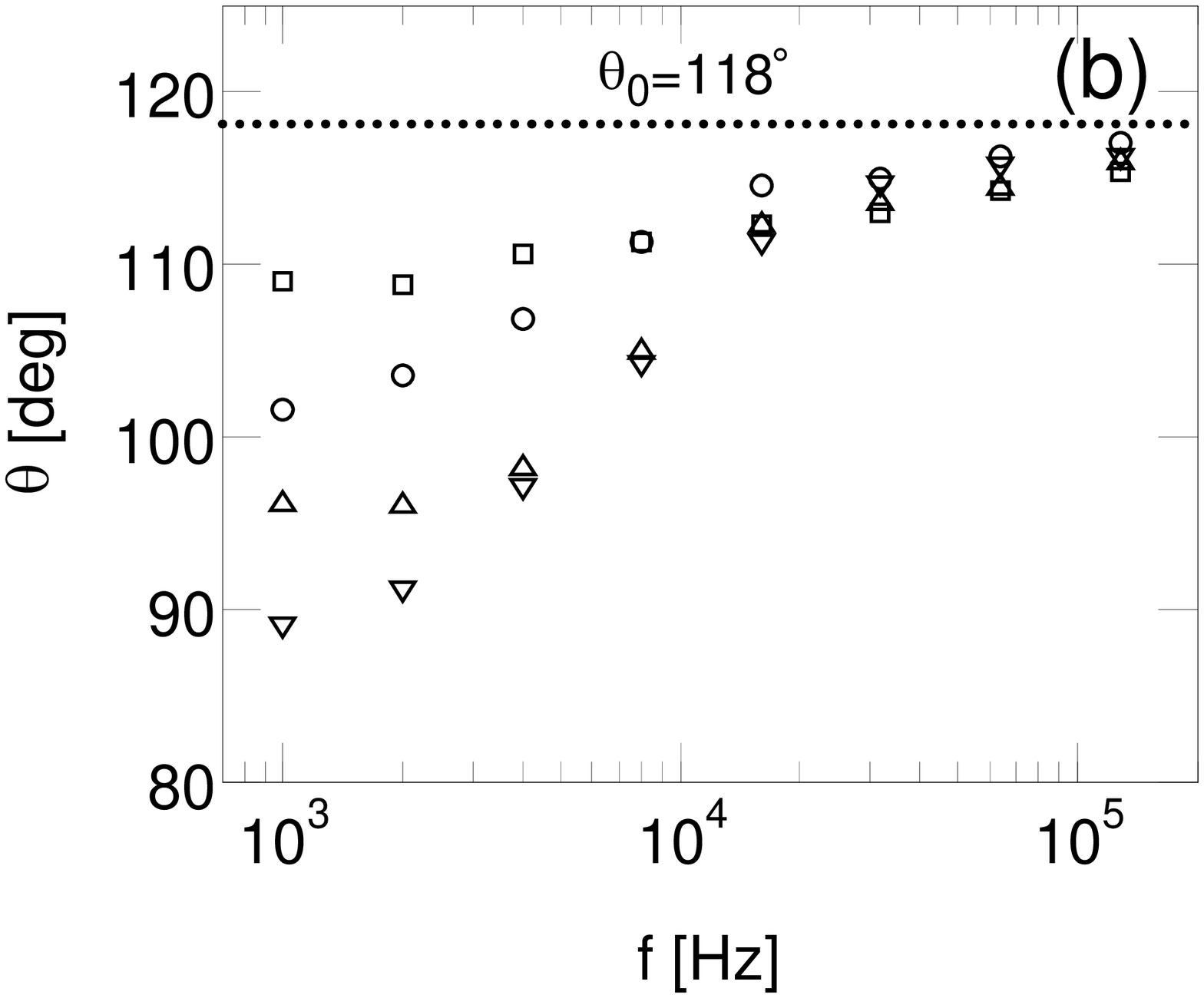}
\caption{\label{fig:freq}\textsf{Contact angle as function of AC frequency $f$ for
several applied voltages $U$. (a) Calculated values corresponding to 57\,V (solid line), 93\,V (short dashed line),
113\,V (long dashed line) and 143\,V (dashed-dotted line) with parameters as in table~\ref{tab:system-freqs}
and $\tau_{b}=0.1\textrm{ms},\,\tau_{r}=2.4\textrm{ ms}$. The value of $\theta_0=118^\circ$ is
indicated by a dotted line. (b) Experimental
results adapted from Ref~\citenum{FreqEW_2007} for the same voltages as in (a): squares (57\,V), circles (93\,V),
triangles (113\,V) and inverted triangles (143\,V).
A quantitative agreement between the calculation and experiment can be
seen. Note also that the electrowetting effect diminishes at high frequencies as predicted
analytically in eq~\ref{eq:F_el_approx}.}}
\end{center}
\end{figure}

The frequency dependence of the contact angle has been experimentally
studied in Ref~\citenum{FreqEW_2007} for several applied voltages.
Figure~\ref{fig:freq} shows a comparison of a minimization of eq~\ref{eq:Z_AC_model} for a range of frequencies,
with experimental data. The calculations have been performed for
system parameters as in table~\ref{tab:system-freqs}, which have been inferred~\cite{ft1}
from Ref~\citenum{FreqEW_2007}, except for $\lambda_{D}=300$\,nm, which
was used as a fitting parameter to the experimental results. This value
corresponds to ionic strength of less than $10^{-6}$\,M and is compatible with de-ionized water used
in Ref~\citenum{FreqEW_2007}. The build-up time was deduced from the featured experimental results to be $\tau_b=0.1$\,ms, and the relaxation time, $\tau_r=2.4$\,ms,
was calculated using eq~\ref{eq:Tr}  with $\rho=1\,\Omega\cdot m^2$.

%%%%%%%%%%%%%%%%%%%%%%%%%%%%%%%%%%%%%%%%%%%%%%%%%%%%%%%%%%%%%%%%%%%%%%%%%%%%%%
%Table 3  DE-IONIZED
%%%%%%%%%%%%%%%%%%%%%%%%%%%%%%%%%%%%%%%%%%%%%%%%%%%%%%%%%%%%%%%%%%%%%%%%%%%%%%
\begin{table}[tp]
\begin{centering}
\begin{tabular}{|c|l|}
\cline{1-2}
~Parameter  & ~Value  \tabularnewline
\cline{1-2}
~$\varepsilon_{l}$  & ~$80$   \tabularnewline
\cline{1-2}
~$\lambda_{D}$  & ~$300\textrm{ nm}$ (fitted)  \tabularnewline
\cline{1-2}
~$V$  & ~$5\textrm{ \ensuremath{\mu}L}$ \tabularnewline
\cline{1-2}
~$d$  & ~$5\textrm{ \ensuremath{\mu}m}$  \tabularnewline
\cline{1-2}
~$\varepsilon_{d}$  & ~$2.67$  \tabularnewline
\cline{1-2}
~$\gamma_{\rm la}$  & ~$72.8\textrm{ mN/m}$\tabularnewline
\cline{1-2}
~$\gamma_{\rm sa}$  & ~$12.7\textrm{ mN/m}$\tabularnewline
\cline{1-2}
~$\gamma_{\rm sl}$  & ~$47\textrm{ mN/m}$\tabularnewline
\cline{1-2}
~$h_g$  & ~$0.7\textrm{ mm}$\tabularnewline
\cline{1-2}
~$b$ & ~$40\textrm{ \ensuremath{\mu}m}$\tabularnewline
\cline{1-2}
\multicolumn{1}{c}{} & \multicolumn{1}{c}{}
\tabularnewline
\end{tabular}
\par\end{centering}
\caption{\label{tab:system-freqs}\textsf{Parameter values of an electrowetting system of
de-ionized water solution on Miyaline-C/Teflon substrate. The parameter values are inferred
from Ref~\citenum{FreqEW_2007}, except for $\lambda_{D}$, which was
fitted to obtain a quantitative agreement with experiments. The resulting value is indeed
compatible with that of de-ionized
water (ionic strength less than $10^{-6}$\,M)~\cite{ft1}.}}
\end{table}
%%%%%%%%%%%%%%%%%%%%%%%%%%%%%%%%%%%%%%%%%%%%%%%%%%%%%%%%%%%%%%%%%%%%%%%%%%%%%%

%%%%%%%%%%%%%%%%%%%%%%%%%%%%%%%%%%%%%%%%%%%%%%%%%%%%%%%%%%%%%%%%%%%%%%%%%%%%%%
%fig10.eps
%%%%%%%%%%%%%%%%%%%%%%%%%%%%%%%%%%%%%%%%%%%%%%%%%%%%%%%%%%%%%%%%%%%%%%%%%%%%%%
%%%%%%%%%%%%%%%%%%%%%%%%%%%%%%%%%%%%%%%%%%%%%%%%%%%%%%%%%%%%%%%%%%%%%%%%%%%%%%

Comparison of the two plots shows that our model reproduces
rather well the frequency
dependence found in experiments.
It can be seen that the electrowetting effect diminishes with rising frequency, and seems to vanish at $f>100$\,kHz
where there is hardly any deviation from the Young angle. This can be easily understood taking into  account that
for $f>10$\,kHz, $1/f$  becomes small
as compared to the double-layer build-up time $\tau_{b}\simeq 0.1$\,ms.
Under those circumstances, ions move too slowly and cannot build considerable
over-concentrations at the electrodes. Thus, at those high frequencies the period-averaged effect
of the double-layers decreases considerably.

The above results imply that it is of value to further explore
the limit of slow electrochemical processes, $\tau_{r}\gg\tau_{b}$,
$\omega\tau_{r}\gg1$, which leads to a simplified expression for the free energy.
Substituting eq~\ref{eq:Z_AC_model}
into eq~\ref{eq:ACmodelE} we  obtain in this limit
\begin{eqnarray}
|Z_{\rm tot}|&\approx &\frac{\sqrt{\tau_b^2+\omega^{-2}}}{C_{\rm tot}(\theta)}\nonumber\\
F_{\rm el}(\theta,U,\omega)&=& -\frac{1}{2\omega |Z_{\rm tot}|}U^2
\approx -\frac{C_{\rm tot}(\theta)}{ 2\sqrt{1+\omega^{2}\tau_{b}^{2}}}U^{2}
\label{eq:F_el_approx}
\end{eqnarray}
where $C_{\rm tot}^{-1}=C_{1}^{-1}+C_{2}^{-1}$ is the total
capacitance of an equivalent system without any electrochemical processes. The
$(1+ \omega^2\tau_b^2)^{-1/2}$ pre-factor in eq~\ref{eq:F_el_approx} depends on the AC frequency
and reflects a diminishing electrowetting effect for rising frequencies.

We note that the DC limit can be obtained  by first assuming no electrochemical processes,
$\tau_{r} \to\infty$ in eq~\ref{eq:Z_AC_model} (leading to eq~\ref{eq:F_el_approx}),
and only then taking the DC limit of $\omega \to 0$ to get
$F_{\rm el}=-\frac{1}{2}C_{\rm tot}U^2$ of eq~\ref{eq:generalized_freeE}. This limit
can be useful in applications where both the substrate and counter-electrode are dielectrically coated.

If electrochemical
processes are not totally excluded, but are just slow (order of seconds,
in accordance with the value calculated in section IV.A), it is expected that
the results obtained in this work will be applicable for that time scale, above which other
mechanisms might take over. Such time-dependent behavior has been observed in Ref~\citenum{ZeroCapillary_2005}.

%%%%%%%%%%%%%%%%%%%%%%%%%%%%%%%%%%%%%%%%%%%%%%%%%%%%%%%%%%
\subsection{Convergence to the Young-Lippmann Formula}
%%%%%%%%%%%%%%%%%%%%%%%%%%%%%%%%%%%%%%%%%%%%%%%%%%%%%%%%%%
In its DC limit ($\omega\rightarrow0$),
equation~\ref{eq:F_el_approx} provides a pathway to establish a relationship
between our model and the Young-Lippmann formula and to show the conditions under which
the two converge. Using eqs~\ref{eq:geometry} and~\ref{eq:C12} (see Appendix for more details), we have
\begin{equation}
C_{\mathrm{tot}}=C_{1} \left[1+\frac{C_{1}}{C_{2}}\right]^{-1}=
C_{1} \left[1+\beta^{-1}
\frac{2-\xi^{3}}{\xi^{2}-l^{-1}{h_g}\xi}\right]^{-1}
\label{eq:Ctot_YL}
\end{equation}
where $l\equiv({3V}/{\pi})^{1/3}$ is a typical drop length,
\begin{equation}
\label{eq:def_beta}
\beta \equiv \frac{6\epsilon_{l}db}{\epsilon_{d}\lambda_{D}l}
\end{equation}
is a dimensionless parameter, and
\begin{equation}
\label{eq:def_xi}
\xi\equiv\left[\frac{1-\cos\theta}{2+\cos\theta}\right]^{1/3}
\end{equation}
is a monotonically increasing function of $0\le \theta \le \pi$.

As long as the second term in the brackets of eq~\ref{eq:Ctot_YL}
is small, our model
agrees with the standard model, $F_{\rm el}=-\frac{1}{2}C_1U^2$, eq~\ref{eq:Ftot1},
with $C_1=C_{\rm ld}$.
For a typical system, as the one presented in table I, the value of
the constant pre-factor is rather small $\beta^{-1}\simeq 0.01$.
Since ${h_g}/{l}\le\xi\le 2^{\frac{1}{3}}$, it
is clear that the quotient can only be large when $\xi\rightarrow h_g/l$ or,
equivalently, when the contact angle becomes small enough. Otherwise,
the second term is negligible and the two models converge.

Note that this view of the validity of the Young-Lippmann formula as
being related to a certain {\it range of the contact angles} is
a departure from the common approach which regards its validity being related to a certain
{\it range of applied voltages}.

By creating this link between the Young-Lippmann formula
 and our model it can be deduced that the proper way of extending the Young-Lippmann formula
(within its validity range) from DC to AC is to replace
$U^2 \to U^{2}_{\rm rms}/\sqrt{1+\omega^2\tau_{b}^2}$. This is exactly the
how the Young-Lippmann formula was scaled (by $g^{\rm eff}$) in section~IV.A.

%%%%%%%%%%%%%%%%%%%%%%%%%%%%%%%%%%%%%%%%%%%%%%%%%%%%%%%%%%%%%
\subsection{The Saturation Angle for Slow Relaxation ($\omega\tau_r\gg1$, $\tau_r\gg\tau_b$)}
%%%%%%%%%%%%%%%%%%%%%%%%%%%%%%%%%%%%%%%%%%%%%%%%%%%%%%%%%%%%%
Within the slow relaxation framework, eq~\ref{eq:F_el_approx}, the minimization of $F_{\rm el}$
(yielding $\theta_{\rm sat}$)  is equivalent to minimizing the total inverse capacitance
$\alpha\equiv 1/C_{\mathrm{tot}}$. Using eqs~\ref{eq:geometry} and~\ref{eq:C12} we obtain
\begin{equation}
\label{eq:inverse_capacitance}
\alpha\left(\xi\right)=\frac{3d}{2\pi\varepsilon_{0}\varepsilon_{d}l^{2}}
\left[\frac{1}{\xi^{-1}-\frac{1}{2}\xi^{2}}+\frac{2\beta^{-1}}
{\xi-{l}^{-1} h_g}\right]
\end{equation}
Minimizing $\alpha\left(\xi\right)$
yields a 6$^{\text{th}}$ order polynomial in $\xi$:
\begin{eqnarray}
 & & \xi^{6}
-2\beta\xi^{5} + 4\beta l^{-1} h_g\xi^{4}\nonumber\\
  &-&2\left(2 + \beta l^{-2}h_g^{2}\right)\xi^{3} - 2\beta \xi^{2}\nonumber\\
 & +&4\beta{l}^{-1}h_g\xi + 2\left(2 -\beta l^{-2}h_g^{2}\right)=0
\end{eqnarray}
It is possible to examine two separate limits for minimizing $\alpha\left(\xi\right)$, leading
to two simple analytical expressions for $\theta_{\rm sat}$.

%%%%%%%%%%%%%%%%%%%%%%%%%%%%%%%%%%%%%%%%%%%%
\subsubsection{Acute saturation angles (large $\beta$)}
%%%%%%%%%%%%%%%%%%%%%%%%%%%%%%%%%%%%%%%%%%%%
If $\xi^{3}\left(\theta_{\rm sat}\right)\ll2$ then, near its minimum,
eq~\ref{eq:inverse_capacitance} reduces to
\begin{equation}
\left. \alpha\right|_{\xi\simeq\xi_{\rm sat}}  \simeq \frac{3d}{2\pi\varepsilon_{0}\varepsilon_{d}l^{2}}
\left[\xi+\frac{2\beta^{-1}}{\xi-{l}^{-1} h_g}\right]
\end{equation}
with a minimum at $\xi=\xi_{\rm sat}$ that satisfies
\begin{equation}
1-2\beta^{-1}\left(\xi_{\rm sat}-
\frac{h_g}{l}\right)^{-2}  =  0
\end{equation}
The solution yields
\begin{equation}
\label{eq:xi_sat}
\xi_{\rm sat}  =  \sqrt{\frac{2}{\beta}}+\frac{h_g}{l}
\end{equation}
and $\theta_{\rm sat}$ can now be obtained
\begin{equation}
\label{eq:cos_theta_sat}
\cos\theta_{\rm sat}  =  \frac{1-2\xi_{\rm sat}^{3}}{1+\xi_{\rm sat}^{3}}
\end{equation}
Inserting typical values from table~\ref{tab:system} to check for self-consistency,
we get $\xi_{\rm sat}^{3}  \approx  0.173\ll 2$ as required. Using eq~\ref{eq:cos_theta_sat}
the saturation angle in this case is calculated to be $\theta_{\rm sat} \simeq 56.1^\circ$, which is not far from the value obtained by a full numerical calculation, $53.3^\circ$. As a rule of thumb we remark that the above
condition, $\xi^{3}_{\rm sat}\ll2$ holds for $\theta_{\rm sat}$  smaller than $\pi/2$,
for which $\xi^{3}\left(\frac{\pi}{2}\right)=0.5\ll 2$.

%%%%%%%%%%%%%%%%%%%%%%%%%%%%%%%%%%%%%%%%%%%%%%%%%%%%
\subsubsection{Large saturation angles (small $\beta$ and  $h_g/l$)}
%%%%%%%%%%%%%%%%%%%%%%%%%%%%%%%%%%%%%%%%%%%%%%%%%
For systems with small $\beta$ the above approximation
should fail, as is apparent from  eq~\ref{eq:xi_sat}.
For such cases, we can use a different approximation assuming that the gap $h_g$ is small enough,
such that
\begin{eqnarray}
\xi_{\rm sat} & \gg & \frac{h_g}{l}
\end{eqnarray}

The saturation angle can then be found from a different approximated form of $\alpha(\xi)$
(eq~\ref{eq:inverse_capacitance}), near its minimum:
\begin{eqnarray}
\left. \alpha\right|_{\xi\simeq\xi_{\rm sat}}   & \simeq & \frac{3d}{2\pi\varepsilon_{0}\varepsilon_{d}l^{2}}
\left[\frac{1}{\xi^{-1}-\frac{1}{2}\xi^{2}}+\frac{2\beta^{-1}}{\xi}\right]\nonumber\\
 & = & \frac{3d}{\pi\varepsilon_{0}\varepsilon_{d}l^{2}}
 \left[\frac{1}{2-\xi^{3}}+\beta^{-1}\right]\xi^{-1}
\end{eqnarray}
Minimizing $\alpha(\xi)$, we get a quadratic equation in the variable $\xi^3$:
\begin{equation}
\xi^{6}-4\left(\beta+1\right)\xi^{3}+
2\left(\beta+2\right)=0
\end{equation}
whose solution is
\begin{equation}
\xi_{\rm sat}^{3} =  2\left(\beta+1\right)\pm\sqrt{4\left(\beta+1\right)^{2}-2\left(\beta+2\right)}
\end{equation}

For small $\beta$ it is possible to further simplify the expression for $\xi_{\rm sat}$ to obtain
\begin{equation}
\xi_{\rm sat}^{3}  \simeq
2\left(1-\sqrt{\frac{3\beta}{2}}\right)
\end{equation}
Checking for self-consistency, the condition holds for small enough
gaps.

%%%%%%%%%%%%%%%%%%%%%%%%%%%%%%%%
\section{Summary and Outlook}
%%%%%%%%%%%%%%%%%%%%%%%%%%%%%%%%%
In this work we propose a novel approach towards electrowetting that, among other results,
can account for contact angle
saturation (CAS) applicable to some electrowetting setups. The model is based on a generalized version of the
free energy accounting for various electric contributions.  The interplay
between the capillary ($F_{\rm cap}$) and electric ($F_{\rm el}$) terms depends on the applied
voltage $U$, because $F_{\rm el} \sim -U^2$. Therefore, when an external voltage is applied it will
drive the system away from
its capillary free energy minimum and towards its electric free energy minimum.

Our approach is distinctly different from other views of electrowetting
that make use of the Young-Lippmann formula. In our model the electric term can exhibit a variety of
dependencies on the contact angle as determined by the exact system geometry. Particularly,
if the electric term $F_{\rm el}$, has a global minimum at a certain contact angle $\theta_{\rm sat}$,
then for high enough voltages this angle also minimizes (asymptotically) the \emph{total} free energy, $F_{\rm tot}$.
Additional increase of the applied voltage does not change the location of the global minimum and the contact
angle saturates at $\theta_{\rm sat}$. We identify exactly this angle with the saturation angle found in
experiments. This very general assumption ($F_{\rm el}$ with a minimum) is all that is needed to show that in the low-voltage limit a Young-Lippmann compatible $\sim U^2$ behavior is expected, while in the
high-voltage limit a $\sim U^{-2}$ saturation should be present. Numerical calculations suggest that combination of these two limiting behaviors approximates rather well the full expression for $\theta(U)$  in the whole voltage range.
in the whole voltage range.
%We note that the free energy can
%be regarded as being a function of any geometrical
%variable and that using the contact angle is just a matter of convenience.

When applying our approach to  EWOD setups, we take two
contributions to $F_{\rm el}$ into account: (i)
the double-layer at the drop/substrate interface; and,
(ii) another double-layer at the drop/counter-electrode
interface. The latter was previously unaccounted for because
it was considered to be
negligible due to geometry, or that its relaxation time was considered
to be very fast. However, we estimate  the relaxation time to be long (on the order of
seconds) and, therefore, the
effect of the counter-electrode double-layer cannot be neglected for AC systems.
Similarly, it cannot be neglected for low voltages (which will exclude electrochemical
processes from taking place at all)~\cite{lowV_2006}, or in DC applications that include
a dielectrically coated counter-electrode.
Using AC circuit analysis we show that
$F_{\rm el}$ indeed has a global minimum that produces the CAS effect.

The value of the
saturation angle as well as the entire electrowetting curve $\theta\left(U\right)$
can be found numerically for any choice of system parameters,
and our specific choice is inspired by the experiments reviewed in Ref~\citenum{B2app_2005}. There is a
qualitative agreement with experimental results, which includes an initial
compliance with the Young-Lippmann formula (scaled correctly), followed by a cross-over to CAS. The values
obtained for the saturation angle,  cross-over voltage
and  saturation voltage are also compatible with  experimental values.

In addition, we  investigated the frequency dependence of electrowetting. It is
shown that the value of the saturation angle is independent of the AC frequency for
a large range of frequencies (1\,kHz to 100\,kHz) for a specific choice of parameters.
A numerical analysis of the
frequency dependence of electrowetting
was conducted for a set of system parameters inferred from Ref~\citenum{FreqEW_2007},
and shows semi-quantitative agreement with the experiment.
These results show that an approximation of the free energy can be justified, such that the entire frequency
dependence is captured in a scaling factor of the applied voltage. It predicts that
the electrowetting effect should diminish with rising frequency, as indeed found in
experiments~\cite{FreqEW_2007}.

In its DC limit our model can converge to the Young-Lippmann formula,
depending on the values of the Young and saturation angles. We use this result to show a novel
way to extend the Young-Lippmann formula from DC to AC systems. We conclude that the
validity of the Young-Lippmann formula is related not to the range of applied voltages, as it
is commonly viewed, but rather to the accessed range of contact angles. In  commonly used
EWOD setups, which are intentionally devised to have as high a Young angle and as low a saturation
angle as possible (so the effect can be more easily measured), our model is compatible with a
compliance to the Young-Lippmann formula at low voltages (and hence high contact angles).
We note that the DC limit of our model can be most useful in DC applications that employ
low voltages and/or include a dielectrically coated counter-electrode.

Our model does not rely on any leakage mechanisms to predict CAS. Nevertheless, we would like to stress
that leakage mechanisms treated in previous works
%%%
[7,22,26,31-33]
can be added. Interestingly, it is conceivable that  a cross-over between inherent CAS
(as in the present model) and CAS originating from leakage mechanisms is responsible for the
time-dependent saturation angle reported in Ref~\citenum{ZeroCapillary_2005}.

The fact that the saturation angle depends on electric parameters whereas the Young angle depends on
the capillary parameters leads to the surprising possibility of {\it reversed electrowetting}. Therefore,
it may be possible to construct a system in which the Young angle is \emph{lower} than the saturation angle.
In such a system the effect of applying an external voltage would be an \emph{increase} in the contact
angle --- in total contradiction with the Young-Lippmann formula that  allows only a decrease in the contact angle. We give an example of a choice of parameters that should yield reversed electrowetting.

Recently, the separate control of the Young and saturation angles was demonstrated in experiments~\cite{Steckl_2010}.
This ability was utilized to construct a set of dye cells~\cite{dye_cell} that are `complementary' in their opposite response
to applied voltage (black-to-white or vice-versa). We believe that further research in this direction will provide ample
opportunity to test for the existence of reversed electrowetting. Finding such evidence would have a potential
impact that can go much beyond our specific model.

We hope that some of the  predictions presented in this paper will be tested in future
experiments in a quantitative fashion, gaining more insight on electrowetting and the CAS phenomenon. For example,
it will be interesting to study how retracting the counter-electrode and, hence, reducing its
contact area $A_2$ affects the saturation angle, as well as coating it with a dielectric material.
Our results suggest that more research into processes taking place at the counter-electrode is needed,
especially with regards to CAS in DC EWOD setups.

%%%%%%%%%%%%%%%%%%%%%%%%%%%%%%%%%%
\section*{Acknowledgments}
%%%%%%%%%%%%%%%%%%%%%%%%%%%%%%%%%%
We thank M. Bazant, D. Ben-Yaakov, B. Berge, T. Blake, H. Diamant, T.B. Jones, M. Maillard, A. Marmur, F. Mugele,
R. Shamai, A. Steckl, U. Steiner, V. Tsionsky and Y. Tsori for many useful discussions and comments. Support from
the Israel Science Foundation (ISF) under grants no. 231/08 and 1109/09, and the US--Israel
Binational Science Foundation (BSF) under grant no. 2006/055 is gratefully acknowledged.

\appendix*
\section{}

It is convenient to express the geometrical parameters and the capacitances in terms of a monotonic function
of the contact angle
\begin{equation}
\label{eq:a1}
\xi(\theta)=\left(\frac{1-\cos\theta}{2+\cos\theta}\right)^{1/3}
\end{equation}
where $l=(3V/\pi)^{1/2}$, derived from the drop volume $V$, is a characteristic length.

The geometrical parameters defined in eq~\ref{eq:geometry} can then be written as
\begin{eqnarray}
\label{eq:a2}
h&=&l\xi(\theta)\nonumber\\
a&=&\sqrt{\frac{1}{3}}l\left(2\xi^{-1}-\xi^2\right)^{1/2}\nonumber\\
A_1&=&\frac{\pi l^2}{3}\left(2\xi^{-1}-\xi^2\right)\nonumber\\
A_2&=&2\pi b h_g\left(l h_g^{-1}\xi-1\right)
\end{eqnarray}

Combining the above expressions with the definitions of the two capacitances, we obtain
\begin{eqnarray}
\label{eq:a3}
C_1&=& \frac{\pi}{3}\frac{\varepsilon_0\varepsilon_d l^2}{d}\left(2\xi^{-1}-\xi^2\right)\nonumber\\
C_2&=& 2\pi b l\frac{\varepsilon_0\varepsilon_l }{\lambda_D}\left(\xi-l^{-1}h_g\right)
\end{eqnarray}
With the use of a dimensionless parameter
\begin{equation}
\beta=\frac{6\varepsilon_l}{\varepsilon_d}\frac{d b}{\lambda_D l}
\end{equation}
the ratio between the two capacitances can finally be expressed as:
\begin{equation}
\frac{C_1}{C_2}=\beta^{-1}\frac{2-\xi^3}{\xi^2-l^{-1}h_g\xi}
\end{equation}

\newpage{}

%%%%%%%%%%%%%%%%%%%%%%%%%%%%%


\begin{thebibliography}{99}
\bibitem{ft3}
The contribution of the counter-electrode to the capillary energy can be neglected here based on
its small dimensions.
\bibitem{ft6}
Note that this
value means that the counter-electrode does not relax momentarily
as is implied by the naive application of the  Young-Lippmann formula to AC EWOD setups, which
suggests $\tau_r\to 0$.
\end{thebibliography}
\end{document}